\documentclass[twocolumn]{aastex62}
\pdfoutput=1
\usepackage{amsmath,amstext,multirow}
\usepackage[T1]{fontenc}
\usepackage{apjfonts,xcolor}
\usepackage[figure,figure*]{hypcap}

\def\be{\begin{eqnarray}}   \def\ee{\end{eqnarray}}
\def\ben{\begin{equation}\begin{aligned}} \def\een{\end{aligned}\end{equation}}
\shorttitle{New constraints from K2}
\shortauthors{Sharma et al.}
\newcommand{\numax}{$\nu_{\rm{max}}$}
\newcommand{\dnu}{$\Delta\nu$}

\begin{document}
\title {The K2-HERMES Survey: Age and Metallicity of the Thick Disc}
\author{Sanjib Sharma}
\affiliation{Sydney Institute for Astronomy, School of Physics, The University of Sydney, NSW 2006, Australia}
\affiliation{ARC Centre of Excellence for All Sky Astrophysics in Three Dimensions (ASTRO-3D)}
\author{Dennis Stello}
\affiliation{School of Physics, University of New South Wales, Sydney, NSW 2052, Australia}
\affiliation{Sydney Institute for Astronomy, School of Physics,
The University of Sydney, NSW 2006, Australia}
\affiliation{Stellar Astrophysics Centre, Department of Physics and Astronomy, Aarhus University, DK-8000 Aarhus C, Denmark}
\affiliation{ARC Centre of Excellence for All Sky Astrophysics in Three Dimensions (ASTRO-3D)}
\author{Joss Bland-Hawthorn}
\affiliation{Sydney Institute for Astronomy, School of Physics, The University of Sydney, NSW 2006, Australia}
\affiliation{ARC Centre of Excellence for All Sky Astrophysics in Three Dimensions (ASTRO-3D)}
\author{Michael R. Hayden}
\affiliation{Sydney Institute for Astronomy, School of Physics, The University of Sydney, NSW 2006, Australia}
\affiliation{ARC Centre of Excellence for All Sky Astrophysics in Three Dimensions (ASTRO-3D)}
\author{Joel C. Zinn}
\affiliation{Department of Astronomy, The Ohio State University, Columbus, OH 43210, USA}
\author{Thomas Kallinger}
\affiliation{Institute of Astrophysics, University of Vienna, Türkenschanzstrasse 17, Vienna 1180, Austria}
\author{Marc Hon}
\affiliation{School of Physics, University of New South Wales, Sydney, NSW 2052, Australia}
\author{Martin Asplund}
\affiliation{Research School of Astronomy \& Astrophysics, Australian National University, ACT 2611, Australia}
\affiliation{ARC Centre of Excellence for All Sky Astrophysics in Three Dimensions (ASTRO-3D)}
\author{Sven Buder}
\affiliation{Max Planck Institute  for Astronomy (MPIA), Koenigstuhl 17, D-69117 Heidelberg}
\affiliation{Fellow of the International Max Planck Research School for Astronomy \& Cosmic Physics at the University of Heidelberg}
\author{Gayandhi M. De Silva}
\affiliation{Department of Physics \& Astronomy, Macquarie University, Sydney, NSW 2109, Australia}
\author{Valentina D'Orazi}
\affiliation{INAF -Osservatorio Astronomico di Padova}
\author{Ken Freeman}
\affiliation{Research School of Astronomy \& Astrophysics, Australian National University, ACT 2611, Australia}
\author{Janez Kos}
\affiliation{Faculty of Mathematics and Physics, University of Ljubljana, Jadranska 19, 1000 Ljubljana, Slovenia}
\author{Geraint F. Lewis}
\affiliation{Sydney Institute for Astronomy, School of Physics, The University of Sydney, NSW 2006, Australia}
\author{Jane Lin}
\affiliation{Research School of Astronomy \& Astrophysics, Australian National University, ACT 2611, Australia}
\author{Karin Lind}
\affiliation{Max Planck Institute  for Astronomy (MPIA), Koenigstuhl 17, D-69117 Heidelberg}
\affiliation{Department of Physics and Astronomy, Uppsala University, Box 516, SE-751 20 Uppsala, Sweden}
\author{Sarah Martell}
\affiliation{School of Physics, University of New South Wales, Sydney, NSW 2052, Australia}
\author{Jeffrey D. Simpson}
\affiliation{School of Physics, University of New South Wales, Sydney, NSW 2052, Australia}
\author{Rob A. Wittenmyer}
\affiliation{Centre for Astrophysics, University of Southern Queensland, Toowoomba, Queensland 4350, Australia}
\author{Daniel B. Zucker}
\affiliation{Department of Physics \& Astronomy, Macquarie University, Sydney, NSW 2109, Australia}
\affiliation{Research Centre in Astronomy, Astrophysics \& Astrophotonics, Macquarie University, Sydney, NSW 2109, Australia}
\author{Tomaz Zwitter}
\affiliation{Faculty of Mathematics and Physics, University of Ljubljana, Jadranska 19, 1000 Ljubljana, Slovenia}
\author{Timothy R. Bedding}
\affiliation{Sydney Institute for Astronomy, School of Physics, The University of Sydney, NSW 2006, Australia}
\author{Boquan Chen}
\affiliation{Sydney Institute for Astronomy, School of Physics, The University of Sydney, NSW 2006, Australia}
\author{Klemen Cotar}
\affiliation{Faculty of Mathematics and Physics, University of Ljubljana, Jadranska 19, 1000 Ljubljana, Slovenia}
\author{James Esdaile}
\affiliation{School of Physics, University of New South Wales, Sydney, NSW 2052, Australia}
\author{Jonathan Horner}
\affiliation{Centre for Astrophysics, University of Southern Queensland, Toowoomba, Queensland 4350, Australia}
\author{Daniel Huber}
\affiliation{Institute for Astronomy, University of Hawai`i, 2680 Woodlawn Drive, Honolulu, HI 96822, USA}
\affiliation{SETI Institute, 189 Bernardo Avenue, Mountain View, CA 94043, USA}
\affiliation{Stellar Astrophysics Centre, Department of Physics and Astronomy, Aarhus University, Ny Munkegade 120, DK-8000 Aarhus C, Denmark}
\author{Prajwal R. Kafle}
\affiliation{International Centre for Radio Astronomy Research (ICRAR), The University of Western Australia, 35 Stirling Highway, \\Crawley, WA 6009, Australia}
\author{Shourya Khanna}
\affiliation{Sydney Institute for Astronomy, School of Physics, The University of Sydney, NSW 2006, Australia}
\author{Tanda Li}
\affiliation{Sydney Institute for Astronomy, School of Physics, The University of Sydney, NSW 2006, Australia}
\author{Yuan-Sen Ting}
\affiliation{Institute for Advanced Study, Princeton, NJ 08540, USA}
\affiliation{Department of Astrophysical Sciences, Princeton University, Princeton, NJ 08544, USA}
\author{David M. Nataf}
\affiliation{Department of Physics and Astronomy, The Johns Hopkins University, Baltimore, MD 21218, USA}
\author{Thomas Nordlander}
\affiliation{Research School of Astronomy \& Astrophysics, Australian National University, ACT 2611, Australia}
\affiliation{ARC Centre of Excellence for All Sky Astrophysics in Three Dimensions (ASTRO-3D)}
\author{Hafiz Saddon}
\affiliation{School of Physics, University of New South Wales, Sydney, NSW 2052, Australia}
\author{Gregor Traven}
\affiliation{Faculty of Mathematics and Physics, University of Ljubljana, Jadranska 19, 1000 Ljubljana, Slovenia}
\author{Duncan Wright}
\affiliation{Centre for Astrophysics, University of Southern Queensland, Toowoomba, Queensland 4350, Australia}
\author{Rosemary F. G. Wyse}
\affiliation{Department of Physics and Astronomy, The Johns Hopkins University, Baltimore, MD 21218, USA
}

\begin{abstract}
Asteroseismology is a promising tool to study Galactic structure and evolution because it can probe the ages of stars. Earlier attempts comparing seismic data from the {\it Kepler} satellite with predictions from Galaxy models found that the models predicted more low-mass stars compared to the observed distribution of masses.
It was unclear if the mismatch
was due to inaccuracies in the Galactic models, or the unknown aspects of the selection function of the stars. Using new data from the K2 mission, which has a well-defined selection function, we find that an old metal-poor thick disc, as used in previous Galactic models, is incompatible with the asteroseismic information. We show that spectroscopic measurements of [Fe/H] and [$\alpha$/Fe] elemental abundances from the GALAH survey indicate
a mean metallicity of $\log (Z/Z_{\odot})=-0.16$ for the thick disc. Here $Z$ is the effective solar-scaled  metallicity, which is a function of [Fe/H] and [$\alpha$/Fe].
With the revised disc metallicities, for the first time, the theoretically predicted distribution of seismic masses show excellent agreement with the observed distribution of masses. This provides an indirect verification of the asteroseismic mass scaling relation is good to within five percent.
Using an importance-sampling framework that takes the selection function into account, we fit a population synthesis model of the Galaxy to the observed seismic and spectroscopic data. Assuming the asteroseismic scaling relations are correct,  we estimate the mean age of the thick disc to be about 10 Gyr, in agreement with the traditional idea of an old $\alpha$-enhanced thick disc.
\end{abstract}

\section{Introduction}
In recent years, asteroseismology has emerged as a powerful tool to study Galactic structure and evolution \citep{2009A&A...503L..21M,2011Sci...332..213C,2013MNRAS.429..423M,2014ApJ...787..110C,2016MNRAS.455..987C,2016ApJ...822...15S,2017ApJ...835..163S,2017A&A...597A..30A,2017MNRAS.467.1433R,2018MNRAS.475.5487S} However, previous attempts based on  data from the original {\it Kepler} mission \citep{2010Sci...327..977B,2013ApJ...765L..41S}, which was designed for detecting transiting planets, have struggled to match the predictions of stellar-population-synthesis Galactic models to observations, with the models producing too many low mass stars \citep{2016ApJ...822...15S,2017ApJ...835..163S}. There are three possible causes for this mismatch:
(i) inaccuracies in the selection function of stars in the observational catalog, (ii) an incorrect Galactic model,
and (iii) systematics in the scaling relations used to relate asteroseismic observables ($\Delta \nu, \nu_{\rm max}$) to density and surface gravity of the stars.
Using data from the K2 Galactic Archaeology Program \citep[K2GAP,][]{2015ApJ...809L...3S,2017ApJ...835...83S}, which is a program to observe oscillating giants with the 'second-life' {\it Kepler} mission \citep[K2,][]{2014PASP..126..398H}
following a well-defined selection function, illuminates the first cause. In this paper we therefore, first use the  K2GAP data to test if  predictions of the Galactic models that are constrained independently of the asteroseismic data match the observed asteroseismic data.
This provides an indirect
way to test the scaling relations.
Having shown that the scaling relations
are fairly accurate, next, we make use of them and the asteroseismic data to fit some of the  parameters in the Galactic model, and discuss the implications for our understanding of the Galaxy.
Unlike {\it Kepler}, however, the seismic detection completeness of K2 is not 100\%. This is because the time span of K2 light curves (typically 80 days) is much shorter  than that of {\it Kepler} (typically more than a year). Hence, we carefully study the detection completeness in K2, and devise ways to take them into account when comparing observations to models.

In a Galactic model, the mass distributions of giants is sensitive to the age and the metallicity of the stellar populations in the model.
While the role of age was investigated in \citet{2016ApJ...822...15S}, the possibility
of an inaccurate prescription of metallicity
being responsible for the mismatch between
observed and predicted mass distributions
has not been investigated so far.
Many studies have attempted to
characterize the metallicity distribution of the thin and thick discs.
For the thin disc, there is a well defined radial metallicity gradient \citep[$\approx$-0.07 dex/kpc][]
{2003A&A...409..523R, 2014AJ....147..116H}
but the age-metallicity relation is almost flat \citep{Bensby2014,2016MNRAS.455..987C,2017ApJS..232....2X,2018MNRAS.475.5487S}. For the thick disc, there is a lack of consensus regarding its properties.  The Besan\c{c}on model adopted a mean metallicity value of [Fe/H]=-0.78 based upon spectroscopic measurements by \citet{1995AJ....109.1095G} (mean thick disc [Fe/H] $\sim -0.6$) and photometric (U, B, V bands) measurements by \citet{1996A&A...305..125R}. The {\sl Galaxia} model \citep{2011ApJ...730....3S}
also adopted the same prescription for metallicity distribution of Galactic components as the Besan\c{c}on model.
However, at least four separate studies have compared predictions of the Besan\c{c}on Galaxy model with that of spectroscopic observations and find that away from the midplane and in regions where the thick disc dominates, the metallicity distribution of the model is inconsistent with observations and that a shift of the thick disc metallicity from -0.78 to about -0.48 is required to make the model agree with  observations. Specifically, \citet{ 2003A&A...398..141S} concluded the thick disc metallicity to be [Fe/H]$=-0.48 \pm 0.05$ by spectroscopically studying about 400 red clump stars in the direction of the North Galactic Pole at a height of $200 < z/{\rm pc} < 800$. Their spectra covered 390-680 nm at $R \sim 42000$. \citet{2011A&A...535A.107K} reached their conclusions by studying a sample of about 700 F, G, and K dwarfs at a height $1 < z/{\rm kpc} < 4$ using the GIRAFFE spectrograph (820.6-940 nm at $R\sim 6500$).
They suggested overall metallicity [M/H] $\sim -0.48$ for the thick disc. Note, the [M/H] of
\citet{2011A&A...535A.107K} is probably close to [Fe/H] but the exact relationship is not known.  \citet{2013A&A...559A..59B} and \citet{2014A&A...568A..71B}
used {\sl Galaxia}  to compare predictions of the Besan\c{c}on model with stars from RAVE (841-879.5 nm at $R\sim 7500$) and  found that for $|z|> 800$ pc, a better match to observations is obtained if [Fe/H] of the thick disc is set to $-0.5$. The former study makes use of dwarfs while the latter uses giants.
More recently, results of \citet{2015ApJ...808..132H} using giants from the APOGEE survey (1.51-1.70 $\mu$m at $R\sim 22500$) also suggest a higher metallicity ([Fe/H$]=-0.36$) for the thick disc, by considering stars between $1 < z/{\rm kpc} < 2$ and Galactocentric radius of $5 < R/{\rm kpc} < 7$ to be thick disc stars.
The TRILEGAL
Galactic model \citep{2005A&A...436..895G}
uses an effective metallicity
of Z=0.008 for the thick disc,
which implicitly takes the $\alpha$ enhancement into account.
This translates to [Fe/H] $\sim -0.5$ (assuming $Z_{\odot}=0.0152$ and [$\alpha$/Fe]=0.24), which compared to the Besan\c{c}on model is
more in line with recent spectroscopic measurements but is still lower than the APOGEE measurements.

We now have a large sample of stars with very precise metallicity measurements from spectroscopic  surveys like the Galactic Archaeology for HERMES \citep[GALAH][]{2015MNRAS.449.2604D} , K2-HERMES \citep[a GALAH-like survey dedicated to K2 follow-up,][]{2018AJ....155...84W} and Apache Point Observatory Galactic
Evolution Experiment \citep[APOGEE][]{2017AJ....154...94M} surveys. In this paper, we use data from the GALAH survey to
determine
the metallicity distribution of the stellar populations in the Besan\c{c}on-based Galactic model that we later use for asteroseismic analysis. Observationally,
it is difficult to measure the metallicity of the stellar populations like the thin and the thick discs that are used in Galactic models.
This is because the thick and thin discs overlap considerably
such that it is difficult to identify individual stars
belonging to each of the discs. Hence, we adopt a forward modeling approach where we fit a Galactic model to the observed data and try to answer the following question.
What is the metallicity of the thick and thin discs that best describes the spectroscopic data from GALAH?
Next, we use data from the APOGEE survey to verify our best fit model.
Unlike spectroscopic studies
of {\it Kepler} seismic targets,
the selection function of the K2-HERMES stars is the same as for the K2 seismic targets.
We take advantage of this fact to then
directly check if the metallicity
distribution of the asterosesimic data in K2, whose mass distributions we wish to compare with Galactic models, is in agreement with the models.

Finally, unlike the original {\it Kepler} survey, which was confined to one direction of the sky, the K2 targets span along the ecliptic allowing us to test our Galactic models in various regions of the Galaxy. Of particular importance is the ability of K2 to investigate the thick disc of the Milky Way. The thick disc is one of the most intriguing components of the Galaxy and its origin is not well understood.  Compared to the thin disc, it is old, alpha-enhanced, metal poor, has higher velocity dispersion, and has a larger scale height. A complication with
the thin disc vs. thick disc nomenclature, is that the scale-length for the thick disc  is shorter than for the thin disc \citep{2012ApJ...753..148B, 2017ApJS..232....2X,2017MNRAS.471.3057M}; the thick disc truncates near the solar circle where the thin disc dominates and beyond the solar circle the thin disc flares with increasing Galactic radius \citep[see discussion and Fig. 1 in][]{blandhawthorn2018}.
Although numerous spectroscopic surveys have targeted thick disc stars, a characterization of the thick disc using asteroseismic data has not been carried out.
It was not possible to do using the {\it Kepler} data because
its field of view was close to the Galactic plane. To move beyond this limitation, a number of K2GAP campaigns were selected at high Galactic latitudes, which means that a significant fraction of stars in the K2GAP are expected to be thick disc stars. Here, we use the K2 data to answer the following question. What is the thick disc age that best describes the seismic masses and spectroscopic data?

The paper is structured as follows.
In Section~\ref{data}, we describe the asteroseismic and spectroscopic data used for the study. In Section~\ref{methods}, we discuss the methods that we use. Here we describe the selection function of the sample and discuss how we take it into account when forward modeling the simulated Galactic data. In Section~\ref{results}, we present our results where we compare model predictions with observations and also tune the metallicity and the age distributions in our model to fit the data. In Section~\ref{conclusions} we discuss and conclude our findings.

\section{Data}\label{data}
\subsection{Target selection}
\begin{figure*}[tb!]
\centering \includegraphics[width=0.8\textwidth]{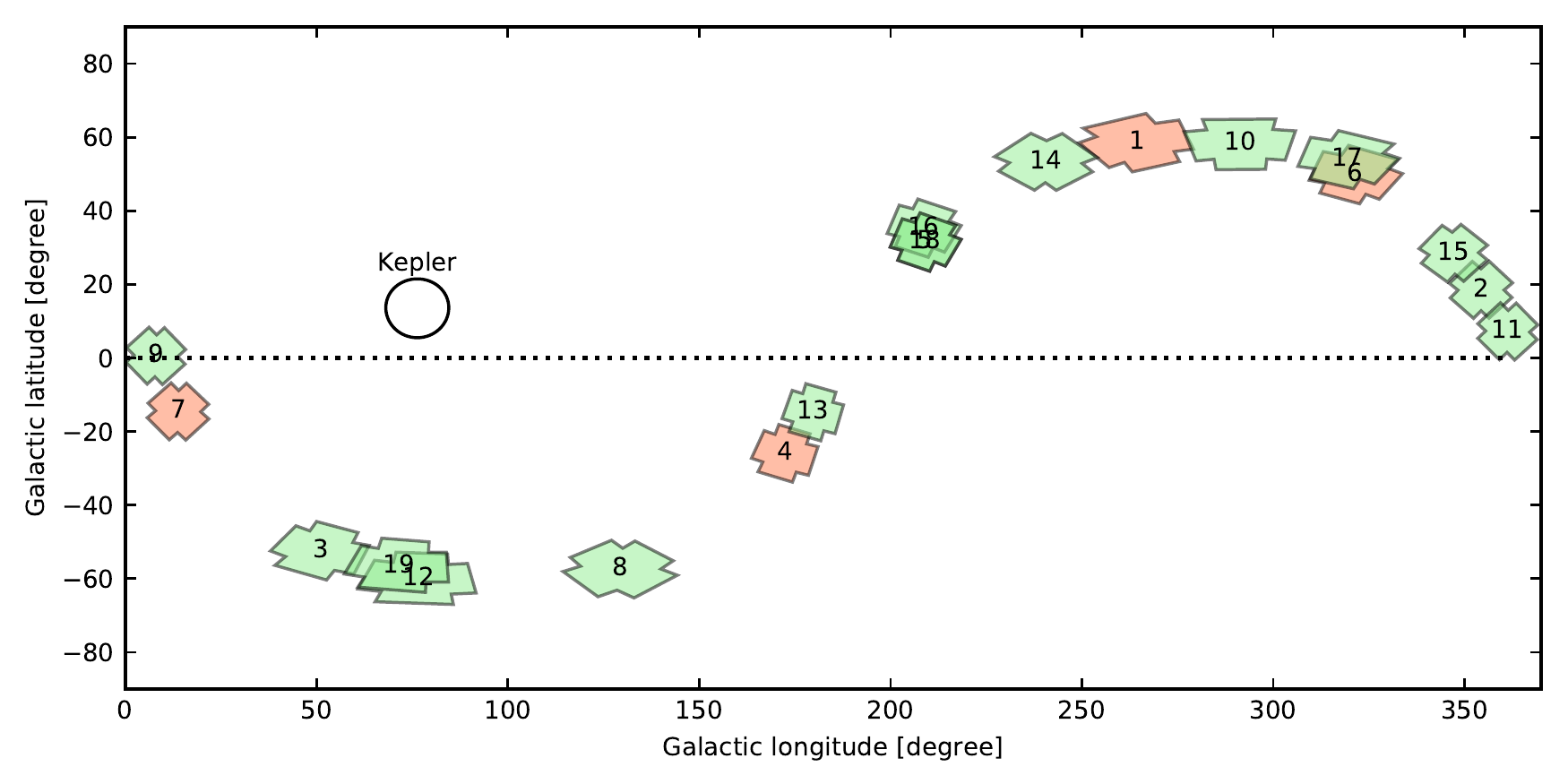}
\caption{Field of view of K2 campaigns 1 to 19 in Galactic coordinates. Orange fields (campaigns C1, C4, C6 and C7)
are studied in this paper. Campaigns C1 and C6 are at high Galactic latitude pointing away from the Galactic midplane, while campaigns C4 and C7 are at low Galactic latitude close to the Galactic plane. Campaign C7 is towards the Galactic center at (0,0) and C4 is towards the Galactic anti-center. \label{fig:fov1}}
\centering \includegraphics[width=0.95\textwidth]{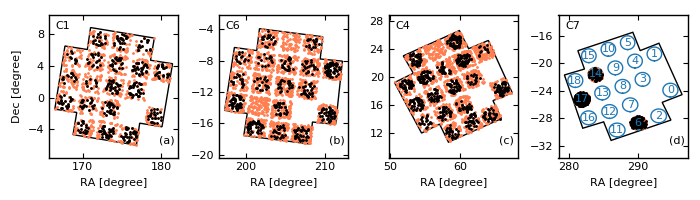}
\caption{
Angular distribution of the seismic sample on sky (orange dots). Each field of view comprises of 21 CCD modules with a small spacing between them. Two CCD modules were broken and hence no observations fall within them.
Black dots are stars with spectroscopic information from the K2-HERMES survey.
The black dots lie within a circle of 1 degree radius centered on the CCD modules.
Campaigns C1, C4, and C6 have no angular selection for the seismic sample. In C7, stars are confined to three circular regions. Panel (d) shows the K2-HERMES sky pointings and their identifier. \label{fig:fov2}}
\end{figure*}

\begin{table*}[tb!]
\caption{Number of K2 GAP targets for each campaign}
\begin{tabular}{@{} l l l l l l l l l}
\hline
Campaign  & Proposed & Observed & Following & N$_{\rm giants}$ & N$_{\rm giants}$ & N$_{\rm giants}$ & Selection function \\
&  &           & selection & with $\nu_{\rm max}$ & with $\Delta\nu$ & $\nu_{\rm max}$ +spec. & \\
\hline
\hline
1  & 9108  & 8630 & 8598 & 1104 & 583  & 455  & $((J-K_s)>0.5)\&(7 < H < 12.927)$ \\
6  & 8371  & 8311 & 8301 & 1951 & 1452 & 504  & $((J-K_s)>0.5)\&(9 < V_{JK} < 15.0)$ \\
4  & 17410 & 6357 & 4937 & 1839 & 945  & 702  & $((J-K_s)>0.5)\&(9 < V_{JK} < 13.447)$ \\
7  & 8698 & 4361 & 4085 & 1541 & 1041 & 930    & $(J-K_s)>0.5)\&$\\
&       &      &      & & & & ($(9 < V_{JK} < 14.5)\&(c \in \{6,17\})$ OR   \\
&       &      &      & & & & ($(14.276 < V_{JK} < 14.5)\&(c \in \{14\})))$  \\\hline
\end{tabular}
\label{tab:tb2}
\tablecomments{Circular pointing identifier $c$ is shown in \autoref{fig:fov2}d.}
\end{table*}
\begin{table}[htb!]
\caption{2MASS quality selection criteria}
\begin{tabular}{@{}llll}
\hline
fFlag  & K2GAP & K2-HERMES & Description \\
& criterion & criterion & \\
\hline
\hline
Qflag & $\leq$ 'BBB' & $=$ 'AAA'    & J,H,K photometric quality\\
Bflag & $=$ '111'    & $=$ '111'    & blend flag \\
Cflag & $=$ '000'    & $=$ '000'    & contamination flag\\
Xflag & $=$ 0        & $=$ 0        & \\
Aflag & $=$ 0        & $=$ 0        & \\
prox  & $>$ 6 arcsec & $>$ 6 arcsec & distance to nearest star\\
\hline
\end{tabular}
\label{tab:tb3}
\end{table}

The stars in this study were observed by K2 as part of the K2GAP Guest Observer program \citep{2015ApJ...809L...3S,2017ApJ...835...83S}. The stars that we use span four K2 campaigns $-$ C1, C4, C6, and C7 $-$ whose sky distributions are shown in \autoref{fig:fov1}.
These K2 campaigns cover different regions of the Galaxy and sample a wide variety of Galactic stellar populations including old, young, thin disc, thick disc, inner disc, and outer disc. C1 and C6 are at high galactic latitudes
and hence are likely to have more old thin disc and thick disc stars owing to the larger scale height of such stars. C4 and C7 are at lower latitudes and are likely to be dominated by young thin disc stars. C4 is towards the Galactic anti-center and samples the outer disc, whereas C7 is towards the Galactic center and samples the inner disc.

The stars follow a simple color magnitude selection based on the 2MASS photometry, which is given in  \autoref{tab:tb2}.
The following equation was used to convert 2MASS magnitude to an approximate
$V$ band magnitude \citep{2018MNRAS.473.2004S}.
\begin{equation}
V_{JK}=K + 2.0 (J-K_s+0.14)+0.382\exp[(J-K-0.2)/0.50]
\end{equation}
Stars having good quality photometry from 2MASS were used; the exact criterion is shown in \autoref{tab:tb3}. Also listed
are the criteria for the spectroscopic sample from the K2-HERMES survey, which for the 'Qflag' is slightly stricter than for the K2GAP targets.
Only a subset of the K2GAP stars observed by K2 were observed by the K2-HERMES survey the number of which is also listed in \autoref{tab:tb2}.
The K2-HERMES survey observes
K2GAP stars that have
magnitudes in range $10<V_{JK}<15.0$ and that lie
in 1 degree radius circular zones.
There are 19 such K2-HERMES zones for each K2 campaign and their layout for campaign C7 is shown in \autoref{fig:fov2}d.
In \autoref{tab:tb2}, $c$ denotes the pointing identifier of these 1 degree radius zones.

In the K2GAP survey, the proposed stars were ranked in priority by $V$ magnitude. For the dense field of C7, they were additionally restricted to only three circular zones, to make the spectroscopic follow-up more efficient. During the final K2 mission-level target selection process for each campaign, the K2GAP target list was truncated at an arbitrary point ($V$-magnitude) based on target allocation. Hence those selected stars will follow the K2GAP selection function. However, targets from other successful Guest Observer programs that overlap with lower ranked (fainter) K2GAP targets could still end up being observed. These stars would not satisfy the K2GAP selection function. It is straightforward to locate the truncation point from the lists of proposed and observed targets by plotting the fraction of proposed to observed stars as a function of row number. A sharp fall in this ratio identifies the location of the truncation point.
The K2GAP-proposed stars, the K2GAP-observed stars, and the K2GAP-observed stars following the K2GAP selection function are listed in \autoref{tab:tb2}. 

\subsection{Asteroseismic data}
Our primary asteroseimic sample comes
from stars observed by K2.
The K2 time-series photometry is sampled roughly every 30 minutes, and span about 80 days per campaign. This allows us to measure the seismic signal in giants brighter than {\it Kepler} magnitude, $Kp$, of $\sim15$ in the range  $10 \lesssim \nu_{\rm max}/\mu{\rm Hz} \lesssim 270$ ($1.9 \lesssim \log g \lesssim 3.2$), with a slight detection bias against the faint high $\log g$ stars due to their higher noise levels and lower oscillation amplitudes \citep{2017ApJ...835...83S}.
We adopt the sesimic results from \citet{2017ApJ...835...83S} (C1) and Zinn et al. (2019,in prep) (C4, C6, and C7). 
Throughout the paper we focus on stars with $10 < \nu_{\rm max}/\mu{\rm Hz}<270$.
A detailed description of the seismic analysis is given in \citet{2017ApJ...835...83S} and reference therein. 
In summary,  automated analysis pipelines perform the measurements of the two seismic quantities used here: the frequency of maximum oscillation power, $\nu_{\rm max}$, and the frequency separation between overtone oscillation modes, $\Delta \nu$. Here, we use the results of the two pipelines called BAM (Zinn et al. 2019 in prep) and CAN \citep{2010A&A...522A...1K}, which are both based on Bayesian MCMC schemes for accessing whether oscillations are detected in a given dataset, and to obtain statistically robust uncertainties on each measurement.
Typically, only 50-70\% of stars for which oscillation are detected (meaning $\nu_{\rm max}$ is determined) do the pipelines also measure a robust $\Delta \nu$  \citep[][Zinn et al. 2019 accepted]{2017ApJ...835...83S}. 

In this paper, in addition to comparing  the predictions of our new Galactic model against results from K2, we also compare against the results from the {\it Kepler} mission. For this
we use the catalog of oscillating giants by \citet{2013ApJ...765L..41S}, in which the global seismic parameters were estimated using the \citet{2009CoAst.160...74H} pipeline (SYD). The exact selection function
of oscillating giants in {\it Kepler} is not known. However, an approximate formula
\be
3.731R_{\odot}<R_{\rm KIC}<\frac{3.7R_{\rm Earth}}{\sqrt{7.1\sigma_{\rm LC}/55.37}},
\ee
was derived by \citet{2016ApJ...822...15S} and we use this to sub-select targets from the above catalog. Here,
$R_{\rm KIC}$ is the photometry-based stellar radius as given in the {\it Kepler} input catalog of \citet{2011AJ....142..112B},
\be
\sigma_{\rm LC}=(1/c_{\rm Kepler})\sqrt{c_{\rm Kepler}+7\times10^6{\rm max}(1,{\it Kp}/14)^4}
\label{equ:jenkins}
\ee
is the long cadence (LC) noise to signal ratio, and $c_{\rm Kepler}=3.46\times10^{0.4(12- Kp)+8}$ is the number of detected $e^{-1}$ per LC sample \citep{2010ApJ...713L.120J}.
For comparing the {\it Kepler}
results with Galactic models,
the synthetic $g$ band SDSS photometry was corrected using Equation 4 from \citep{2016ApJ...822...15S} and then
$R_{\rm KIC}$ was estimated from synthetic photometery using the procedure outlined in \citet{2011AJ....142..112B}.

\subsection{Spectroscopic data}
The spectroscopic data come from the K2-HERMES (for seismic K2 targets) and the GALAH surveys (non seismic targets) being conducted at the 3.9-m AAT located at Siding Spring observatory in Australia. The spectra were collected using the multi object High Efficiency and Resolution Multi-Element Spectrograph (HERMES) spectrograph  \citep{2015JATIS...1c5002S}.
The K2-HERMES survey uses the same instrument setup as the GALAH survey \citep{2017MNRAS.465.3203M} and the TESS-HERMES survey \citep{2018MNRAS.473.2004S}.
The reduction is done using a custom IRAF based pipeline \citep{2017MNRAS.464.1259K}. The spectroscopic analysis is done using
the GALAH pipeline and is described in \citep{2018MNRAS.478.4513B}. It uses
Spectroscopy Made Easy (SME) to first build a training set by means of a model driven scheme \citep{2017A&A...597A..16P}. Next, {\it The Cannon} \citep{2015ApJ...808...16N} is used to estimate the stellar parameters and abundances by means
of a data driven scheme.

\section{Methods}\label{methods}
\subsection{Galactic models}
In this paper we perform two kinds of analysis, one is to compare the predictions of theory with observations and the other is to fit Galactic models to the observed data.
For this, we use population synthesis based Galactic models. The models consist of four different Galactic components, the thin disc, the thick disc, the bulge, and the stellar halo. The full distribution of stars in space, age, and metallicity $Z$, is given by
\be
p(R,z,Z,\tau|\theta)=\sum_k p(k) p(R,z,Z,\tau|\theta,k),
\ee
with $k$ denoting a Galactic component, $\theta$ the parameters governing the Galactic model,
$R$ the Galactocentric cylindrical radius, $z$ the Galactic height, $\tau$ the age, and $Z$ the metallicity of the stars in the Galactic component.
To sample data from a prescribed
population synthesis model we use the {\sl Galaxia}\footnote{\url{http://galaxia.sourceforge.net}} code
\citep{2011ApJ...730....3S}. It uses a Galactic model
that is initially based on the {\sl Besan\c{c}on} model
by \citet{2003A&A...409..523R} but with some crucial modifications. The density laws and the initial mass functions for the various components are given in
Table 1 of \citet{2011ApJ...730....3S} and are same as in \citet{2003A&A...409..523R}.
The density normalizations  for various components are given in \autoref{tab:gmodels1}, these differ slightly from \citet{2003A&A...409..523R}
and a discussion of the changes is given in Section 3.6 of \citet{2011ApJ...730....3S}.
The thin disc spans an age range of 0 to 10 Gyr and has a star formation rate which is almost constant. The thin disc has a scale height which increases with age according to Equation 18 in \citet{2011ApJ...730....3S}. In this paper, we leave the thick disc
normalization as a free parameter and solve for it using data from Gaia DR2.

Other differences between {\sl Galaxia} and the {\sl Besan\c{c}on} model are as follows. {\sl Galaxia} is a robust statistical sampler, it provides continuous sampling over any arbitrary volume of the Galaxy. This enables rigorous comparisons with observed stellar surveys for an arbitrary selection function.
{\sl Galaxia} has a 3D extinction scheme that is based on
\citet{1998ApJ...500..525S} dust maps.
We also apply a low-latitude correction to the
dust maps as described in \citet{2014ApJ...793...51S}.
The isochrones used to predict the stellar
properties are from the Padova database
using CMD 3.0 (\href{http://stev.oapd.inaf.it/cmd}{http://stev.oapd.inaf.it/cmd}), with  PARSEC-v1.2S isochrones \citep{2012MNRAS.427..127B, 2014MNRAS.445.4287T, 2014MNRAS.444.2525C, 2015MNRAS.452.1068C}, the NBC version of bolometric corrections \citep{2014MNRAS.444.2525C}, and assuming Reimers mass loss with efficiency $\eta=0.2$ for RGB stars. The isochrones are computed for scaled-solar composition following the $Y=0.2485+1.78Z$ relation and their solar metal content is $Z_{\odot}=0.0152$.
\begin{table}[tb!]
\caption{Galactic models with different age and metallicity distribution functions.}
\begin{tabular}{@{} l l l l}
\hline
Model& & Thick & Thin\tablenotemark{a}  \\
\hline
MP\tablenotemark{b}&$\langle{\rm [M/H]}\rangle$   & -0.78 & [0.01, 0.03, 0.03,  0.01,\\
& & &  -0.07, -0.14, -0.37] \\
& $\sigma_{\rm [M/H]}$ & 0.33  & [0.12,    0.12,    0.10,  0.11, 0.18,\\
& & & 0.17,  0.2] \\
& Min(Age\tablenotemark{c})& 11& [0, 0.15, 1, 2, 3, 5, 7] \\
& Max(Age)& 11& [0.15, 1, 2, 3, 5, 7, 10] \\
\hline
MR&$\langle{\rm [M/H]}\rangle$   & -0.16 & [0.01, 0.03, 0.03,  0.01, 0, 0, 0] \\
& $\sigma_{\rm [M/H]}$&  0.17 & same as MP \\
& Min(Age)& 9& same as MP \\
& Max(Age)& 11& same as MP \\
\hline
FL&$\langle{\rm [M/H]}\rangle$   & -0.14 & 0.0 \\
& $\sigma_{\rm [M/H]}$ &  0.30  & 0.3 \\
& Min(Age)& 6& same as MP \\
& Max(Age)& 13& same as MP \\
\hline
\end{tabular}
\label{tab:gmodels}
\tablenotemark{a}{Thin disc consists of 7 distinct populations with different age ranges and ${\rm d [M/H]/d} R=-0.07$ dex/kpc} \\
\tablenotemark{b}{The [M/H] values correspond to [Fe/H] values used by  \citet{2003A&A...409..523R}, ignoring $\alpha$ enhancement.} \\
\tablenotemark{c}{In units of Gyr}
\end{table}

The details of the different Galactic models that we use in this paper are given in \autoref{tab:gmodels}. The base {\sl Galaxia} model denoted by MP (metal poor) is from \citet{2011ApJ...730....3S}, it has an old metal poor thick disc and a thin disc whose mean metallicity decreases with age as in \citet{2003A&A...409..523R}. The model denoted by MR (metal rich) has metal rich thick and thin discs. The FL (flat) model also has a metal rich thick and thin disc, but unlike other models its thick disc spans an age range from 6 to 13 Gyr with a uniform star formation rate and no variation of metallicity with age. For each Galactic component $k$, the IMF, the formula for spatial distribution of stars, and the density normalizations are given in \citet{2011ApJ...730....3S}.

\begin{table}
\caption{The IMFs and the density normalizations of Galactic components. The parameters $\alpha_1$ and t$\alpha_2$ are used to specify the IMF (number density of stars as a function of mass stellar mass $M$), which is of the following form, $\propto M^{\alpha_1}$ for $M/{\rm M}_{\odot}<1$ and $\propto M^{\alpha_2}$ for $M/{\rm M}_{\odot} >1$.}
\begin{tabular}{l l l l}
\hline
Galactic Component & Normalization & $\alpha_1$  & $\alpha_2$\\
\hline
Thin ($0<{\rm Age/Gyr}<7$)& \tablenotemark{a}2.37 ${\rm M_{\odot} yr^{-1}}$ & -1.6 & -3.0 \\
Thin ($7<{\rm Age/Gyr}<10$)& \tablenotemark{a}1.896 ${\rm M_{\odot} yr^{-1}}$ & -1.6 & -3.0 \\
Thick & \tablenotemark{b}$\rho_{\rm \odot,thick}$ & -0.5 & -0.5 \\
Stellar Halo & \tablenotemark{b}$10.252\times 10^{3} {\rm M_{\odot}pc^{-3}}$ & -0.5 & -0.5 \\
Bulge & \tablenotemark{c}13.76 ${\rm stars\ pc^{-3}}$ & -2.35 & -2.35 \\
\hline
\end{tabular}
\label{tab:gmodels1}
\tablenotemark{a}{Star formation rate for an IMF spanning a mass range  of 0.07 to 100 $M_{\odot}$.} \\
\tablenotemark{b}{Local mass density of visible stars} \\
\tablenotemark{c}{Central density} \\
\end{table}

To compare predictions of Galactic models to
asteroseismic data, we need to estimate the
observed seismic quantities $\nu_{\rm max}$ and $\Delta \nu$ for the synthetic stars. The seismic quantities are  estimated from effective temperature $T_{\rm eff}$, surface gravity $g$, and density $\rho$ using the following asteroseismic scaling relations \citep{1991ApJ...368..599B,1995A&A...293...87K,1986ApJ...306L..37U}.
\be
\frac{\nu_{\rm max}}{\nu_{\rm max,\odot}}=\frac{g}{g_{\odot}}\left(\frac{T_{\rm eff}}{T_{\rm eff,\odot}}\right)^{-0.5} \label{equ:scaling_numax}\\
\frac{\Delta \nu}{\Delta \nu_{\odot}}=f_{\Delta \nu}\left(\frac{\rho}{\rho_{\odot}}\right)^{0.5} \label{equ:scaling_dnu}
\ee
Here,
\be
f_{\Delta \nu}=\left(\frac{\Delta \nu}{135.1\ \mu{\rm Hz}}\right)\left(\frac{\rho}{\rho_{\odot}}\right)^{-0.5}
\ee
is the correction factor
derived by \citet{2016ApJ...822...15S} by analyzing theoretical oscillation frequencies with GYRE \citep{2013MNRAS.435.3406T} for stellar
models generated with MESA \citep{2011ApJS..192....3P,2013ApJS..208....4P}. We used the code ASFGRID\footnote{\href{http://www.physics.usyd.edu.au/k2gap/Asfgrid}{http://www.physics.usyd.edu.au/k2gap/Asfgrid}} \citep{2016ApJ...822...15S}
that computes the correction factor as a function of metallicity $Z$, initial
mass $M$, evolutionary state $E_{\rm state}$ (pre or post helium ignition), $T_{\rm eff}$, and $g$.

\subsection{Importance-sampling framework}\label{sec:impsamp}
To constrain the parameters of a Galactic model from the observed data we developed and used an importance-sampling framework, which we now describe. Suppose we have collected some data regarding some variable $x$, such as metallicity Z or seismic mass, subject to some selection function $S$. Then suppose that we have a Galactic model parameterized by $\theta$ from which we can draw samples subject to the same selection function $S$. To constrain the model, we start with a base model parameterized by some $\theta_0$, then to change the model to one parameterized by a new $\theta$, we simply reweight the samples from the simulation parametrized  by $\theta_0$ instead of drawing from a new simulation. When the model changes from $\theta_0$ to $\theta$, the new
weights for a star $i$ belonging to a Galactic component $k$ are given by
\be
w_i=p( R_i,z_i,Z_i,\tau_i|\theta,k)/p(R_i,z_i,Z_i,\tau_i|\theta_0,k).
\ee
In general, such a change can alter the number of visible stars of your synthetic Galaxy, but as long as the
parameters governing the density distribution
of the stars are unaltered, the changes are minimal.
In this paper, we are mainly concerned with only
altering the thick disc parameters like mean age and metallicity. We also alter the metallicity of the old thin disc, but this change is minor and can be ignored for the present discussion related to the number of visible stars.
The base model that we use is based on the Besan\c{c}on model, which was constructed by \citet{2003A&A...409..523R} to satisfy the observed star counts in the Galaxy. When the thick disc parameters, like mean age and/or the metallicity are modified,
we adopt the following procedure to address the slight change that is expected in the number of visible thick disc stars.
We measure $f_{\rm SGP}$, the ratio of stars
that lie between $2<|z|/{\rm kpc}<3$
out of  all  stars that are in a $30^{\circ}$ radius cone around the south Galactic pole ($b=-90.0^{\circ}$) and have Gaia magnitudes $0<G<14$ from Gaia DR2 \citep{2018A&A...616A...1G}. Using $f_{\rm SGP}$ estimated from Gaia DR2, we solve for the  normalization factor $\rho_{\rm \odot, thick}$ and reweight the thick disc of the model such that $f_{\rm SGP}$  in the selection-function-matched mock sample matches with that of the Gaia DR2 data. Following this global normalization, the stars are further reweighted to satisfy the color magnitude selection function of the observational data to which the model is being fitted.

To fit the model to the data we need to compute the likelihood of the data given the model and this is done as follows.
Let $x_q$ be the $q_{\rm th}$ percentile of the distribution of some variable $x$. For this variable, suppose we have observed samples $X_{o}$ and samples from some model $X_m$, with the model being parameterized by $\theta$ and $S$ being the selection function. The probability of the observed data given the model can then be written as
\be
p(X_o|\theta,S)=\prod_q \frac{1}{\sqrt{2 \pi} \sigma_x} \exp\left(-\frac{(x_{o,q}-x_{m,q})^2}{2\sigma_x^2}\right), {\rm \ with\ } \label{equ:likelihood}
\\ \sigma_x^2=(\sigma_{x,{\rm o}}^2/n_{\rm eff,o}+\sigma_{x,{\rm m}}^2/n_{\rm eff,m})
\nonumber
\ee
Here, $n_{\rm eff}$ is the effective number of stars, which for stars with different weights is given by $\left(\sum w_i\right)^2/\sum w_i^2$ according to Kish's formula. We make use of 16, 50 and 84 percentiles to compute the likelihood of the data given the model. The $x_{o,q}$ and $x_{m,q}$
denote the $q_{\rm th}$ percentile obtained from samples
$X_o$ and $X_m$ respectively.
For multiple data sets, $X = \{X_{o}^{1}, ... , X_{o}^{ns}\}$ with each of them having their own selection function $S = \{S_1, ... , S_{ns}\}$, the full likelihood is
\be
p(X|\theta,S) = \prod_i p(X_{o}^{i}|\theta,S_i).
\ee

In this paper, the importance-sampling framework is used for estimating the metallicity of the thick disc using spectroscopic data from the GALAH survey and to estimate the age of the thick disc from the asteroseismic data from K2. For the former (metallicity estimation), we bin up the stars lying in $5<R/{\rm kpc}<11$ and $1<|z|/{\rm kpc}<3$ using bin sizes of 0.5 kpc in $R$ and 0.33 kpc in $|z|$. We use $\log Z/Z_{\odot}$ as the observed variable $x$ and fit for the mean and the dispersion of the thick disc metallicity and the mean metallicity of the old thin disc (age greater than 3 Gyr) in the Galactic model. For this we use the MR model from \autoref{tab:gmodels}.
For the latter (age estimation), we bin up the stars into different K2 campaigns and 3 different giant classes.
We follow \citet{2016ApJ...822...15S} by using the temperature-independent seismic mass proxy
\be
\label{equ:scaling_m1}
\kappa_{M} &  = &  \left(\frac{\nu_{\rm
max}}{\nu_{\rm max,
\odot}}\right)^{3}\left(\frac{\Delta \nu}{
\Delta \nu_{\odot}}\right)^{-4}  \\
\ee
as the variable $x$ and fit for the age (mean) and metallicity (mean and dispersion) of the thick disc. For this we use the FL model from \autoref{tab:gmodels}.
For each selection of stars the likelihood is computed using \autoref{equ:likelihood}.
The $\kappa_M$ is closely related to the stellar mass $M$, which is given by
\be
\label{equ:scaling_m}
\frac{M}{\rm {\rm M}_\odot} &  = &  \kappa_{M}
\left(\frac{T_{\rm eff}}{T_{\rm
eff, \odot}}\right)^{1.5}.
\ee
Given that temperatures are not always readily available for the observed stars, we use $\kappa_M$ instead of mass when comparing theoretical predictions to observations. This also removes any ambiguity in temperature scale differences between the models and the data.
For simplicity we will in the following refer to $\kappa_M$ as mass.

\subsection{Detection completeness}
Before we can compare the mass distributions, we have to make sure that the observed and simulated data satisfy the same selection function. In other words, we have to properly forward-model the simulated data and make it satisfy the same observational constraints that the observed data satisfies.

\begin{figure}[tb]
\centering \includegraphics[width=0.48\textwidth]{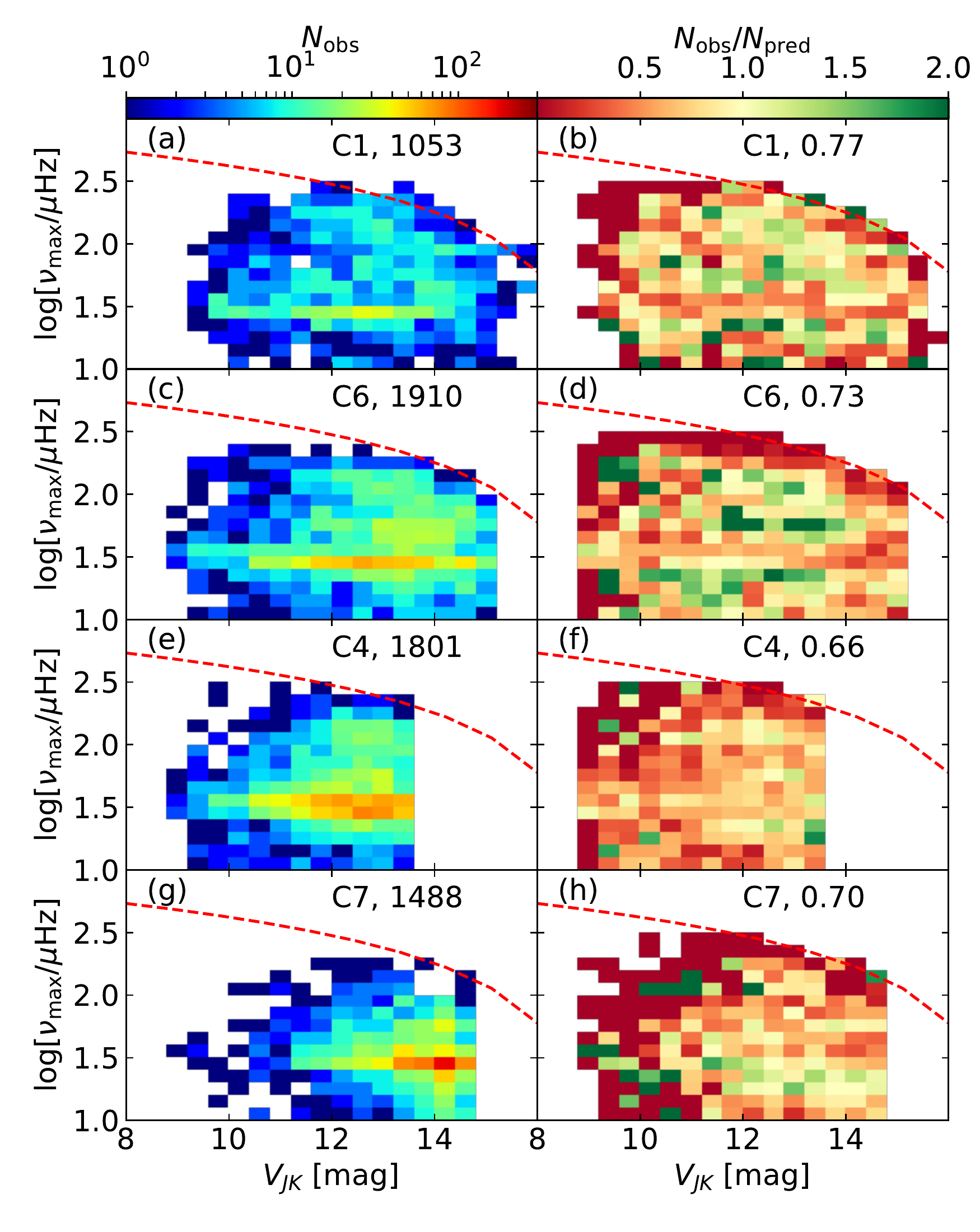}
\caption{Distribution of observed (left panels) stars in the $(\nu_{\rm max},V_{JK})$ plane for four K2 campaigns and the {\it Kepler} field.
The right panels plot the ratio of observed to predicted oscillating giants in each bin. The predictions are based on simulations using {\sl Galaxia}.
The dashed line represents the equation $\nu_{\rm max}=-60(V_{JK}-17)$.
The upper right region (above the dashed line) indicates where we cannot detect oscillations due to too low signal-to-noise.
\label{fig:vmag_numax}}
\end{figure}

The duration of the K2 campaigns sets a lower limit on the detectable $\nu_{\rm max}$ of about $10 \mu$Hz below which the seismic detection efficientcy drops. The observational cadence sets an upper limit of about $\nu_{\rm max}=270 \mu$Hz \citep{2015ApJ...809L...3S}.
The amplitude of oscillations decreases with increasing $\nu_{\rm max}$ (less luminous stars) and the photometric noise increases towards fainter stars. This makes it harder to detect oscillations for stars that have higher $\nu_{\rm max}$  and/or are faint. This bias is clearly visible as missing stars in the top right corner of \autoref{fig:vmag_numax}(a,c,e,g), which shows the distribution of observed stars in the $(\nu_{\rm max},V_{JK})$ plane.

To model the seismic detection probability we followed the scheme presented by \citet{2011ApJ...732...54C} and \citet{2016ApJ...830..138C}. For this, we used the mass, radius, and effective temperature of each synthetic star to predict its total mean oscillation power and granulation noise in the power spectrum.
The oscillation amplitude was estimated as
$A=2.5\left(L/{\rm L}_{\odot}\right)^{0.9}\left(M/{\rm M}_{\odot}\right)^{-1.7}\left(T_{\rm eff}/{\rm T_{eff,\odot}}\right)^{-2}$ following \citet{2011ApJ...737L..10S}.
The granulation power was estimated using the \citet{2014A&A...570A..41K} model.
The apparent magnitude was used to compute the instrumental photon-limited noise in the power spectrum, which combined with granulation noise gave the total noise.
For the instrumental noise we use \autoref{equ:jenkins} \citep[formula given by][]{2010ApJ...713L.120J}. For K2, we scaled the noise by a factor of three to take into account the higher noise in the K2 data compared to the {\it Kepler} data and also applied a minimum threshold of 80 ppm.
The mean oscillation power and the total noise were then used to derive the probability of detecting oscillations, $p_{\rm detect}$, with less than 1\% possibility of false alarm.  Stars with $p_{\rm detect}> 0.9$ were assumed to be detectable.

\begin{figure}[tb]
\centering \includegraphics[width=0.48\textwidth]{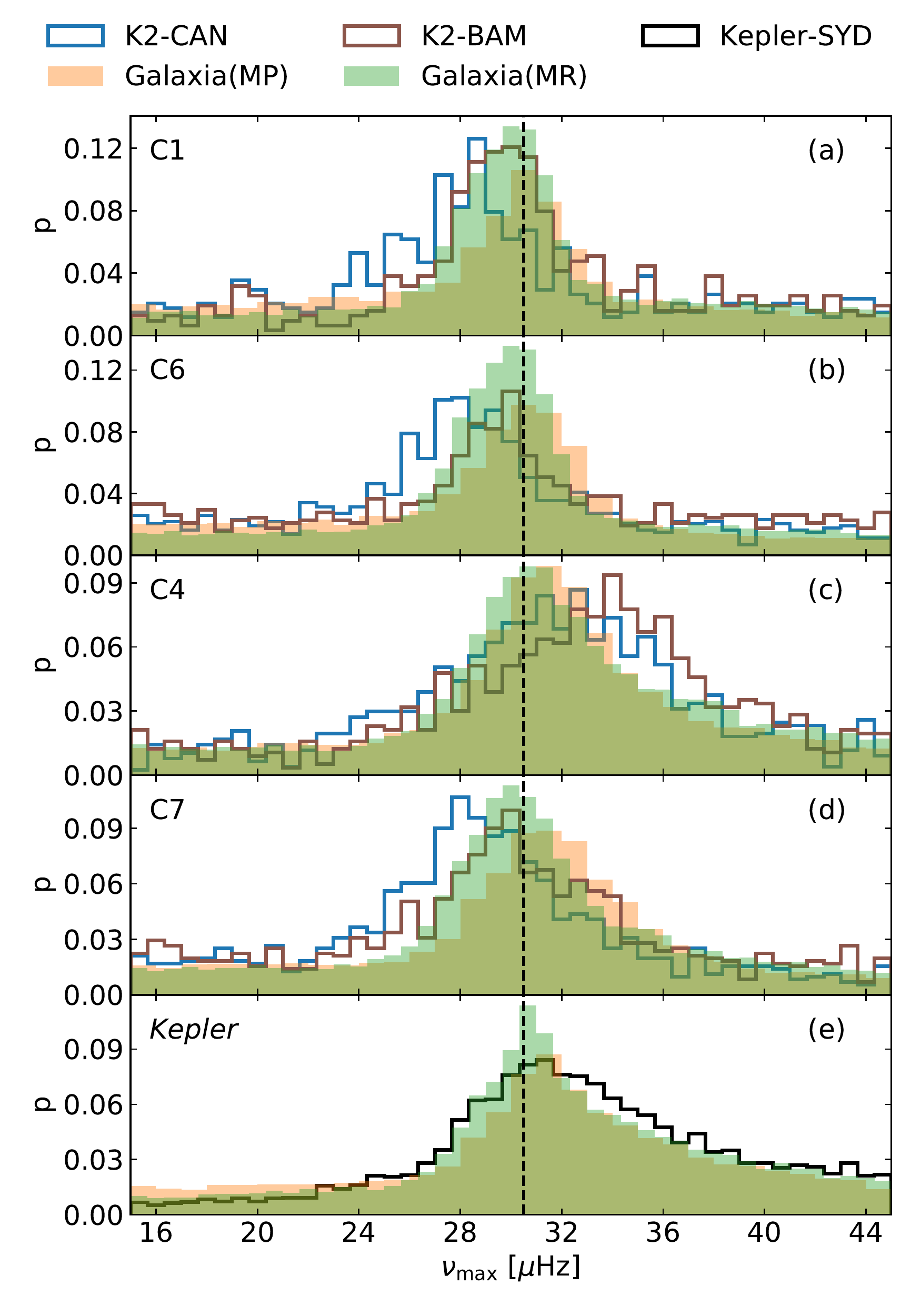}
\caption{The probability distribution of $\nu_{\rm max}$ for observed and predicted oscillating giants.  The dashed line, $\nu_{\rm max}=30.5\ \mu$Hz, shows the approximate location of the peak in the distribution of the predicted stars. The peak corresponds to the location of the red clump giants. The location of the peak is not sensitive to the choice of the Galactic model, but the distribution is sharper for the MR model.
The peak for the CAN pipeline
is systematically lower as compared to the BAM pipeline.
For C4 the location of the peak for both the CAN and the BAM pipelines is higher as compared to predictions.
\label{fig:numax2_dist}}
\end{figure}

The results of applying the detection probability on the {\sl Galaxia} simulated stars are shown in \autoref{fig:vmag_numax}(b,d,f,h) as the ratio between the number of observed to predicted stars.  The $\nu_{\rm max}$ was estimated using \autoref{equ:scaling_numax}.
The figure shows that the fraction of predicted to observed stars is close to one over most of the regions where we have observed stars.
However, C4, C6, and C7 show a slight tendency of having a lower than predicted number of stars towards the top right corner of each panel (higher $V_{JK}$ and higher $\nu_{\rm max}$), where the signal-to-noise ratio of the oscillations is low. This is probably because for these campaigns, the detections are based solely on automated pipelines, with no additional visual inspection as for C1 \citep{2017ApJ...835...83S}. The mean detection fraction is 0.72, and the cause for fewer detections is not clear. Using the  deep-learning-based pipeline of  \citet{2018MNRAS.476.3233H} resulted in slightly more $\nu_{\rm max}$ detections, raising the mean detection fraction to 0.78, but the fraction still remained significantly less than one.

\subsection{The distribution of $\nu_{\rm max}$ and apparent magnitude}
We now check the distribution of apparent magnitudes and $\nu_{\rm max}$ in more detail.
In \autoref{fig:numax2_dist}, the $\nu_{\rm max}$ distributions show a peak, which corresponds to the red clump stars. For the simulated data (orange line) the peak is close to $30.5 \mu$Hz.
The location of the peak varies very little across different campaigns, it is about 1 $\mu$Hz higher for the low latitude campaigns C4 and {\it Kepler}. For the observed data
analyzed with the BAM pipeline,
except for C4, the location of the peak does not show any obvious shift with respect to the predicted peak.
For C4 the BAM peak is about 3 $\mu$Hz higher.
For all campaigns, the location of the peak for the
CAN pipeline is systematically lower by 2 $\mu$Hz compared to the BAM pipeline. This suggests that the CAN pipeline systematically underestimates  $\nu_{\rm max}$ for stars around the red clump region  compared to the scaling relation prediction. The peak for the {\it Kepler} data obtained using the SYD pipeline also did not show any shift with respect to the predicted peak.
To conclude, we see systematic differences between campaigns and between pipelines, they are small
but could be important for certain applications
and hence should be investigated further in future.

\begin{figure}[tb]
\centering \includegraphics[width=0.48\textwidth]{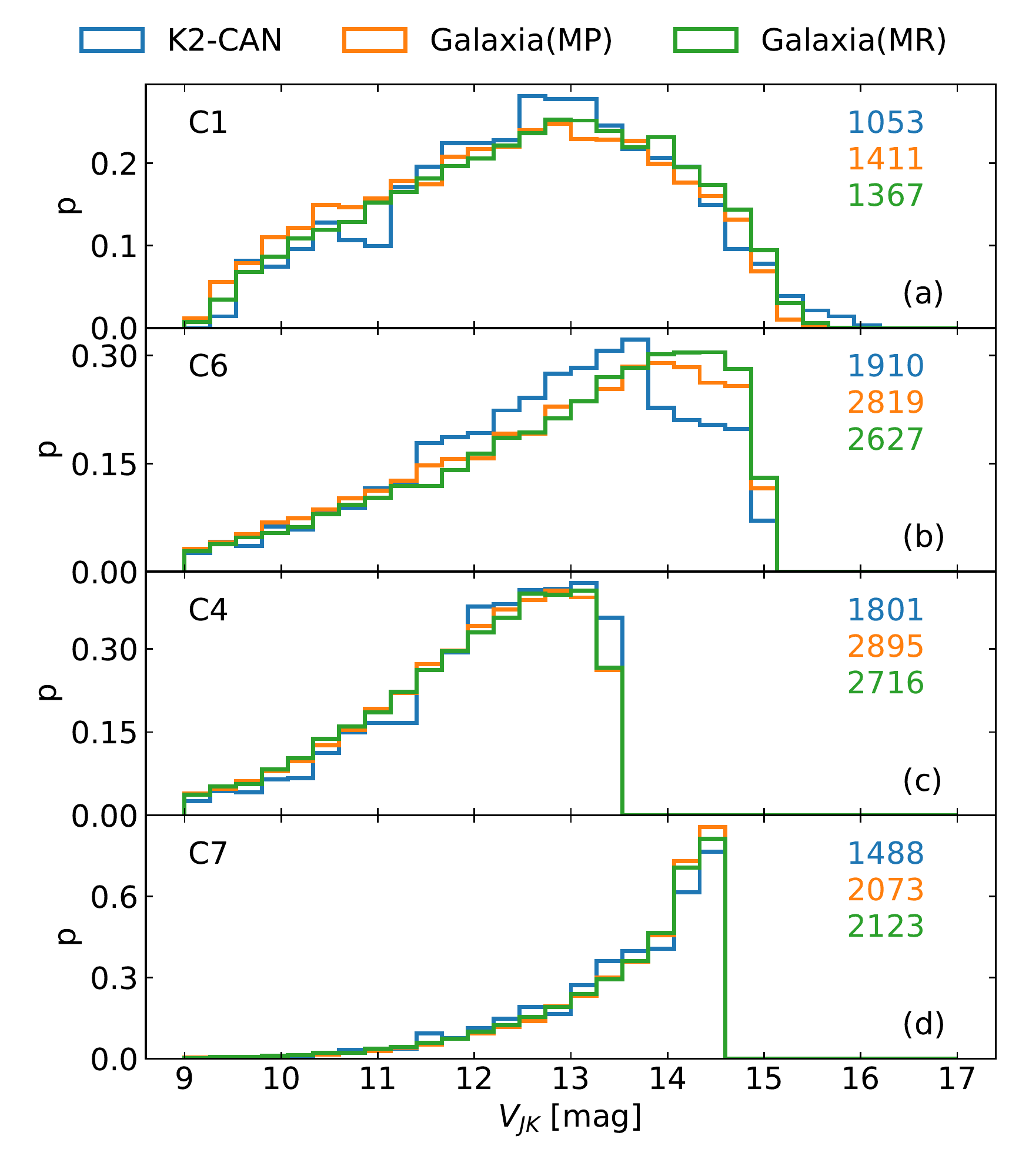}
\caption{
Magnitude distribution of observed oscillating giants from K2 along with predictions from {\sl Galaxia} corresponding to model MP and MR. The number of stars with $\nu_{\rm max}$ detections in the observed sample and those predicted by model MP, and MR are also listed in each panel.}
\label{fig:vmag_dist}
\end{figure}

The corresponding distributions of $V_{JK}$ for K2 are shown in \autoref{fig:vmag_dist}.
Overall the observed distributions match well with the model predictions. For C6, the model
predicts more stars for $V_{JK}>13.5$, the cause of which is not yet clear, but we found that this has no impact on our conclusions related to the mass distribution of stars that we present in this paper.

\subsection{Classifying giants into different classes}\label{sec:giant_class}
In the seismic analysis, $\nu_{\rm max}$ is easier to detect
as compared to $\Delta \nu$.  Hence, there are
stars with a $\nu_{\rm max}$ measurement but no $\Delta \nu$ measurement and this needs to be taken into account when comparing model predictions with observations. To accomplish this, we first study the $\Delta \nu$-detection completeness  of our sample and then devise ways to account for it when comparing model  predictions with observations.

\begin{figure}[tb]
\centering \includegraphics[width=0.48\textwidth]{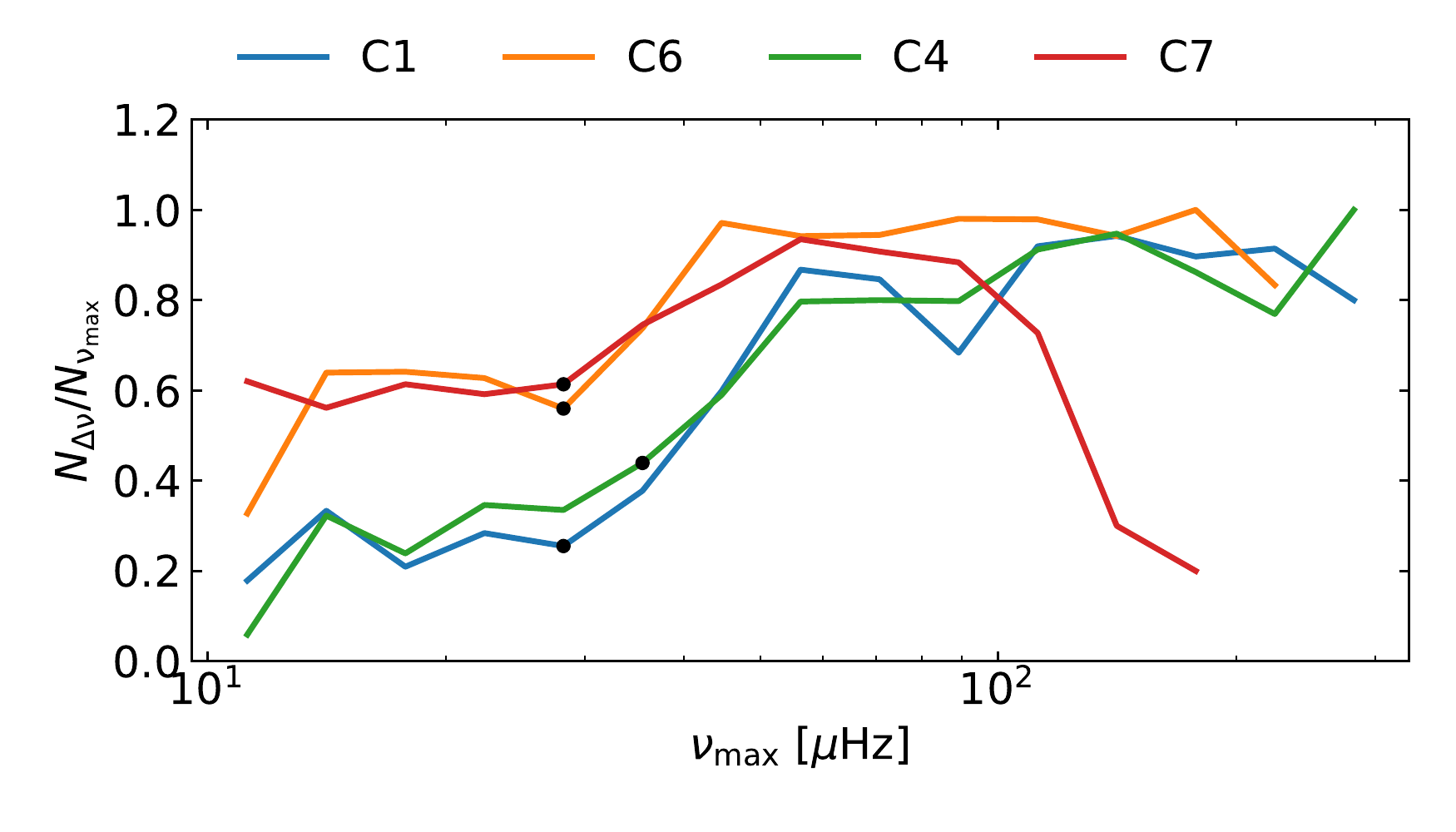}
\caption{The ratio of the number of stars with and without $\Delta \nu$ measurements for various K2 campaigns. The black dot marks the frequency of the peak, $\nu_{\rm max,RC}$, in the $\nu_{\rm max}$ distribution of the observed stars and is due to RC stars. The ratio shows a sharp increase for stars with $\nu_{\rm max}>\nu_{\rm max,RC}$.
\label{fig:numax1_ratio_dnu}}
\end{figure}

\begin{figure}[tb]
\centering \includegraphics[width=0.48\textwidth]{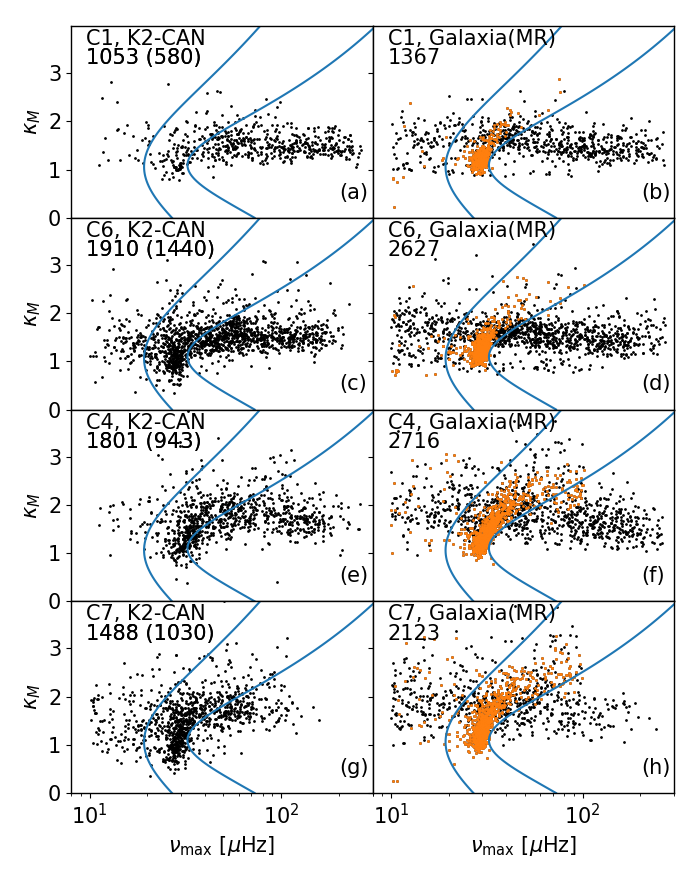}
\caption{ Distribution of stars in the $(\kappa_M, \nu_{\rm max})$ plane.
The blue lines split the plane into three distinct regions, the predominantly high-luminous RGB stars (left), the predominantly red-clump stars (middle) and the low-luminosity RGB stars (right). Left panels (a,c,e,g) show results from K2 based on the CAN pipeline. Right panels (b,d,f,h) show predictions from {\sl Galaxia}. The overplotted orange points denote the red clump stars.
\label{fig:numax_kappa_m}}
\end{figure}

The probability, $p_{\Delta \nu}$, of having a $\Delta \nu$ measurement given that we have a measurement of $\nu_{\rm max}$ is shown in \autoref{fig:numax1_ratio_dnu}, as a function of $\nu_{\rm max}$. This was derived by binning the stars in $\nu_{\rm max}$ and then computing in each bin the ratio of the number of stars with a $\Delta \nu$ measurement ($N_{\Delta \nu}$) to those with a $\nu_{\rm max}$ measurement ($N_{\nu_{\rm max}}$). We see three distinct phases.  The first is for $\nu_{\rm max}<25 \mu$Hz, where $p_{\Delta \nu}$ is constant but low. The second is for $25<\nu_{\rm max}<50 \mu$Hz, where $p_{\Delta \nu}$ increases with $\nu_{\rm max}$. And the third  is for $\nu_{\rm max}>50 \mu$Hz, where $p_{\Delta \nu}$ is again constant and close to 1 (except for C7 where $p_{\Delta \nu}$ is lower for $\nu_{\rm max}>100 \mu$Hz).
The drop in $p_{\Delta \nu}$ as $\nu_{\rm max}$ decreases from 50 to 30 $\mu$Hz, coincides with the increase in fraction of red-clump stars as predicted by a {\sl Galaxia} simulation (see orange dots in \autoref{fig:numax_kappa_m}). This drop could be because the power spectra of red-clump stars are more complex than RGB stars \citep{2013ARA&A..51..353C} and this makes the $\Delta \nu$ measurement harder to obtain.
For $\nu_{\rm max}< 25 \mu$Hz, we mainly have RGB stars, but the $p_{\Delta \nu}$ is still low, and this could be due to the limited frequency resolution of the K2 data starting to affect our ability to obtain a clear $\Delta \nu$ measurement towards the low $\nu_{\rm max}$ stars.
Although all four campaigns show similar $p_{\Delta \nu}$ for high $\nu_{\rm max}$ stars, we note that for $\nu_{\rm max}<30 \mu{\rm Hz}$, $p_{\Delta \nu}$ is about a factor of two lower for C1 and C4 compared to C6 and C7. The cause for this different behavior is not clear.

We have seen in \autoref{fig:numax1_ratio_dnu} that $\Delta \nu$ detections are incomplete with a completeness that depends on stellar type (evolution stage).
This suggests that
we should study the different types of giants separately. Below we describe a scheme to segregate stars into three giant classes, the high luminosity RGB stars, the RC stars, and the low luminosity RGB stars. The segregation is done in the $(\kappa_M, \nu_{\rm max})$ plane.
By construction the high luminosity RGB class will have some contamination from AGB stars and the RC class will have contamination from RGB stars.

\autoref{fig:numax_kappa_m} shows the distribution of stars in the $(\kappa_M, \nu_{\rm max})$ plane both for the observed and ${\sl Galaxia}$-simulated data. Although RC stars typically have $\nu_{\rm max} \sim 30 \mu{\rm Hz}$, \autoref{fig:numax_kappa_m} shows that the high $\kappa_M$ stars can have $\nu_{\rm max}$ reaching up to $\sim 100 \mu{\rm Hz}$.
This is the main reason why we decided not to isolate RGB stars solely from their $\nu_{\rm max}$.
Based on simulations by ${\sl Galaxia}$ we instead fit and obtain two curves
\be
\nu_{\rm max}^{\rm lower} & = & 6.8478\ \kappa_{M}^2 -14.489 \ \kappa_{M}+26.914 \\
\nu_{\rm max}^{\rm upper} & = & 33.598\ \kappa_{M}^2 -73.523 \ \kappa_{M} +72.647
\ee
that enclose about 92\% of the RC stars (blue lines).
In \autoref{fig:numax_kappa_m}, it can be seen that the red clump stars are nicely enclosed by the blue lines.

These curves are then used to classify stars into the
three categories; a) $\nu_{\rm max}<\nu_{\rm max}^{\rm lower}$ (high-luminosity RGB stars or hRGB), b) $\nu_{\rm max}^{\rm lower}<\nu_{\rm max}<\nu_{\rm max}^{\rm upper}$) (RC stars),  and c) $\nu_{\rm max}>\nu_{\rm max}^{\rm upper}$ (low-luminosity RGB stars or lRGB). Based on {\sl Galalxia} simulations, the fraction of RGB stars in the three categories averaged across all campaigns was found to be a) 0.87, b) 0.18, and c) 0.97, suggesting that each category is dominated by the desired stellar type in that category, i.e., RGB, RC and RGB respectively.

\section{Results}\label{results}
\subsection{Constraints from spectroscopic surveys}
Large scale surveys of the Milky Way were not available at the time the {\sl Besan\c{c}on} model was constructed as implemented in {\sl Galaxia}.
The situation has changed now, with surveys like APOGEE and GALAH providing high-resolution spectra for hundreds of thousands of stars, which sample the Galaxy well beyond the solar neighborhood. Hence it is possible to characterize the the thick disc better than before.

To study the elemental composition of the thick disc
we need to identify stars belonging to the thick disc. This can be done using height above the Galactic plane or rotational velocity. We choose the former approach as the overlap of the thin and thick disc is quite strong in rotational velocity. To isolate thick disc stars, we select stars with $(5<R/{\rm kpc}<7)$ and $1<|z|/{\rm kpc}<2$. This region provides the largest number of thick disc stars with the least amount of contamination from the thin disc, as can be seen in Figure 4 from \citet{2015ApJ...808..132H}.
\begin{figure}[tb]
\centering \includegraphics[width=\columnwidth]{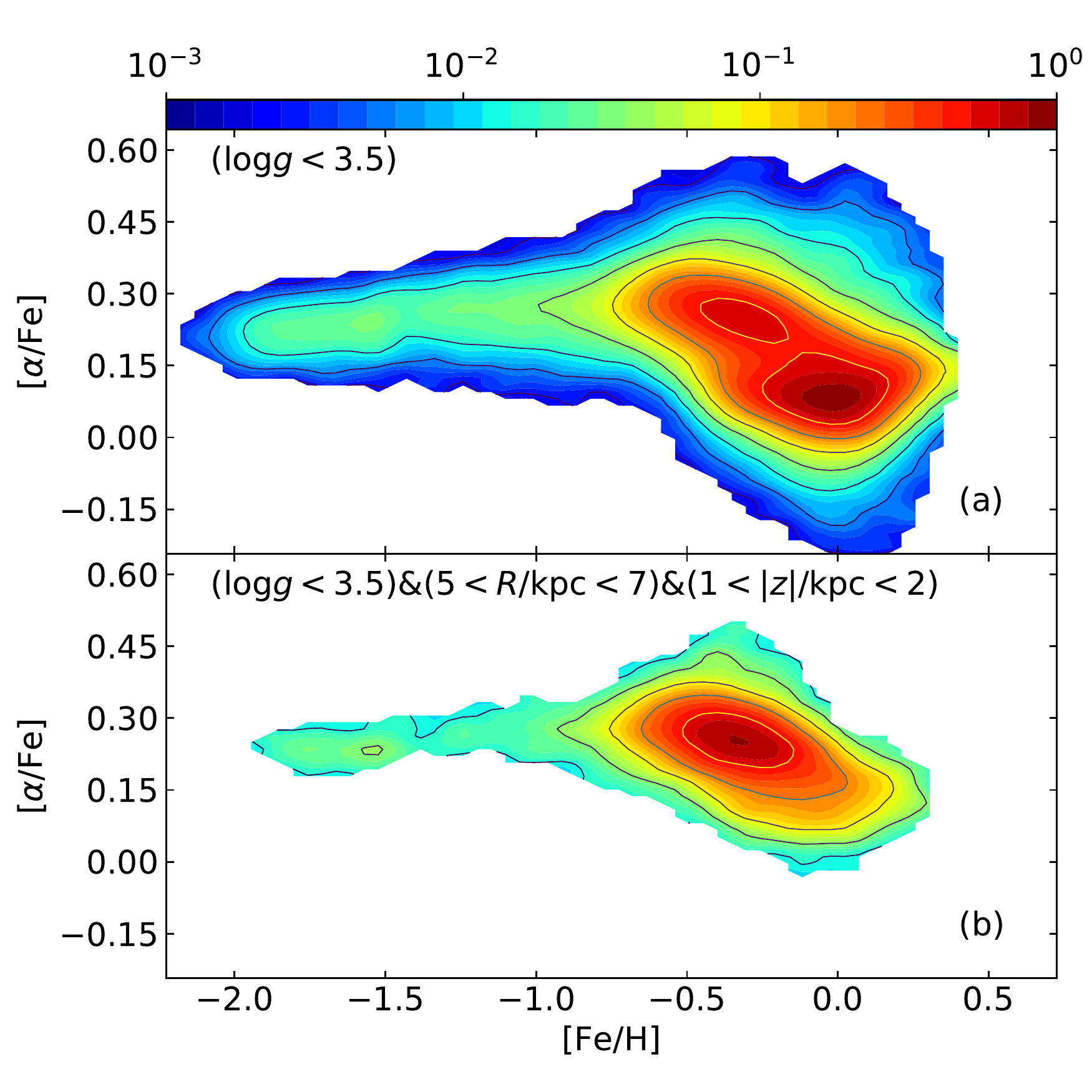}
\caption{Distribution of GALAH giants in the ([Fe/H],[$\alpha$/Fe]) plane.  Giants were selected using $\log g < 3.5$. (a) Distribution of all giants. (b) Giants restricted to $5<R/{\rm kpc}<7$ and $1<|z|/{\rm kpc}<2$. The color bar shows  probability density which is normalized such that the maximum density is 1.
\label{fig:alpha_feh_dist}}
\end{figure}
In \autoref{fig:alpha_feh_dist}, we further illustrate this using data from the GALAH survey \citep{2018arXiv180406041B}. Three populations are visible in \autoref{fig:alpha_feh_dist}a; the stellar halo at [Fe/H]$\sim -1.75$, the thin disc at
[Fe/H]$\sim 0$, and the thick disc at [Fe/H]$\sim -0.39$. After selecting stars by location, the thin disc sequence almost vanishes and the halo can be identified as a separate over-density in  \autoref{fig:alpha_feh_dist}b.
\begin{table}[htb!]
\caption{Abundance of iron and alpha elements for thick disc stars. Median and standard deviation based on 16-th and and 84-th percentile values are listed. The first four rows show the abundances for stars with $5<R/{\rm kpc}<7$ and $1<|z|/{\rm kpc}<2$ and positive rotation about the Galaxy. The last row shows the result obtained by fitting a Galactic model.}
\begin{tabular}{@{}lllllll}
\hline
Source  & \multicolumn{2}{c}{[Fe/H]} & \multicolumn{2}{c}{[$\alpha$/Fe]} & \multicolumn{2}{c}{$\log(Z/Z_{\odot})$}\\
& med & sdev & med & sdev &med & sdev\\
\hline
APOGEE & -0.294 & 0.28 & 0.186 & 0.08 & -0.160 & 0.24 \\
GALAH DR2 & -0.367 & 0.24 & 0.218 & 0.08 & -0.196 & 0.21 \\
GALAH DR2c\tablenotemark{a} & -0.316& 0.21 & 0.239 & 0.07 & -0.131 & 0.18
\\
GALAH mock\tablenotemark{b}& & & & & -0.170 & 0.25 \\
GALAH DR2c\tablenotemark{c} &  & & & & -0.162 & 0.17
\\
\hline
\end{tabular}
\label{tab:al_feh}
\tablenotemark{a}{Calibrated} \\
\tablenotemark{b}{Mock catalog generated by {\sl Galaxia} with $\log(Z/Z_{\odot}) \sim \mathcal{N}(-0.18,0.22^2)$ for the thick disc} \\
\tablenotemark{c}{Fitting a Galactic model to GALAH stars with $5<R/{\rm kpc}<11$, $1<|z|/{\rm kpc}<3$, $12<V_{JK}<14$, \texttt{field\_id}<6546} and positive rotation.
\end{table}
For this particular spatial selection, the median and the spread of the distribution of [Fe/H] and [$\alpha$/Fe] are listed in \autoref{tab:al_feh} and compared with that of APOGEE.
Also given are metallicity estimates [M/H] constructed using the formula \begin{eqnarray}
{\rm [M/H]} & = &
\log\left(\frac{Z}{Z_{\odot}}\right) \nonumber \\
& = & {\rm [Fe/H]}
+ \log(10^{[\alpha/{\rm Fe}]} 0.694+ 0.306).
\end{eqnarray}
by \citet{2005essp.book.....S}.
Given an isochrone grid constructed for metallicities $Z$ using solar-scaled composition with a specified $Z_{\odot}$, the above formula provides an approximate estimate of metallicity $Z$ or [M/H] for a given [Fe/H] and [$\alpha$/Fe].
In \autoref{tab:al_feh}, although we choose to show the median, the mean values were also
very similar with the difference between the two being less than 0.01 dex (after discarding stars with [Fe/H] $<-1.25$, which most likely belong to the stellar halo).

\begin{figure}[tb]
\centering \includegraphics[width=0.48\textwidth]{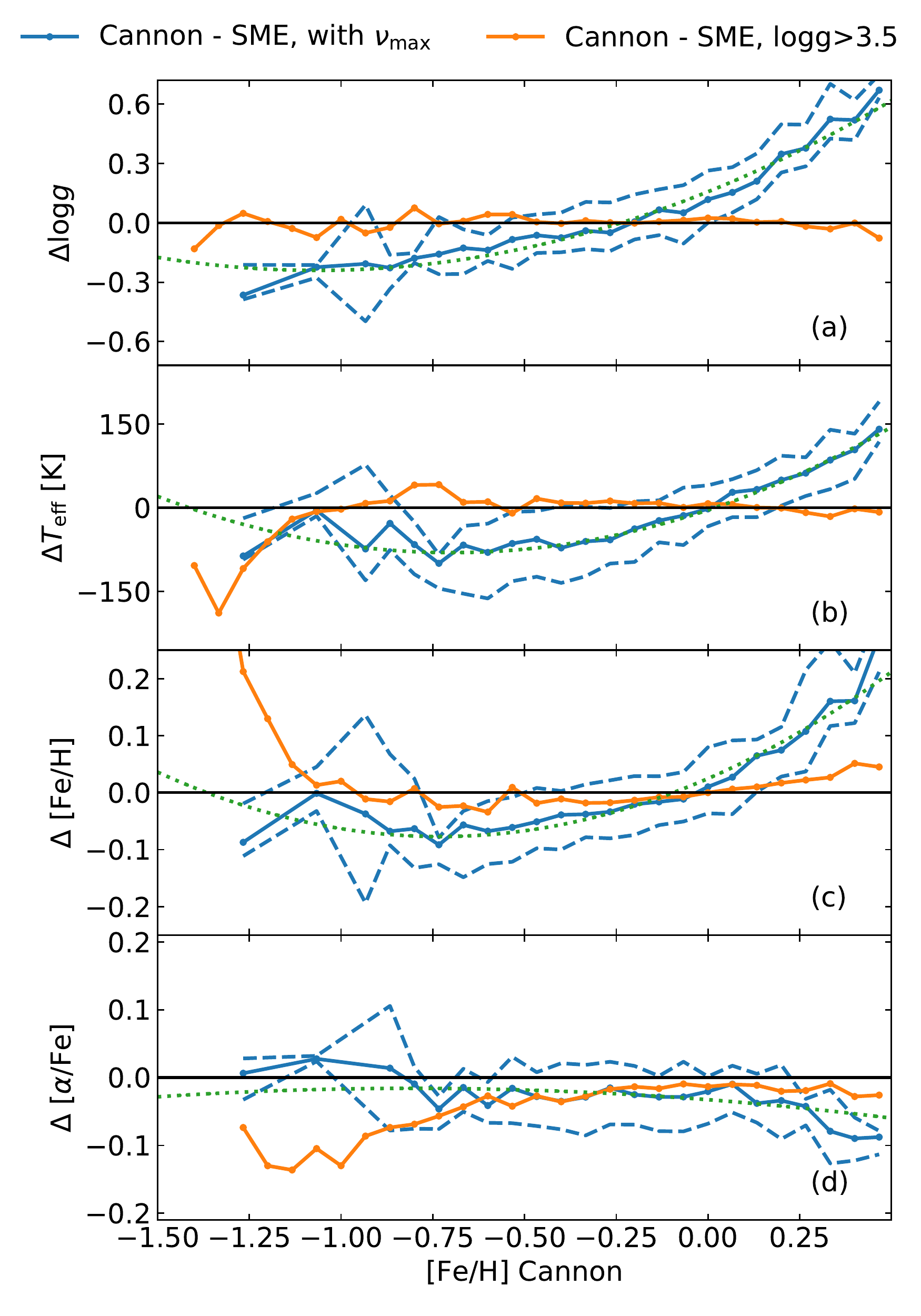}
\caption{Comparison of GALAH-DR2 Cannon-based (data-driven) estimates to that of SME-based (model-driven) estimates. The plots shows systematic trends as a function
of Cannon-based iron abundance [Fe/H]. The giants
(blue) and dwarfs (orange) are shown separately.
The giants shown are seismic giants from K2,
and for them SME was run using
$\nu_{\rm max}$ estimated from asteroseismology as
a prior.
The seismic giants show strong systematic trends while dwarfs have negligible systematics. The dotted line is a two degree polynomial fit to the trends for the seismic giants with $-1.5<{\rm [Fe/H]}<0.3$.
\label{fig:galah_vs_seism}}
\end{figure}

\begin{figure}[tb]
\centering \includegraphics[width=0.48\textwidth]{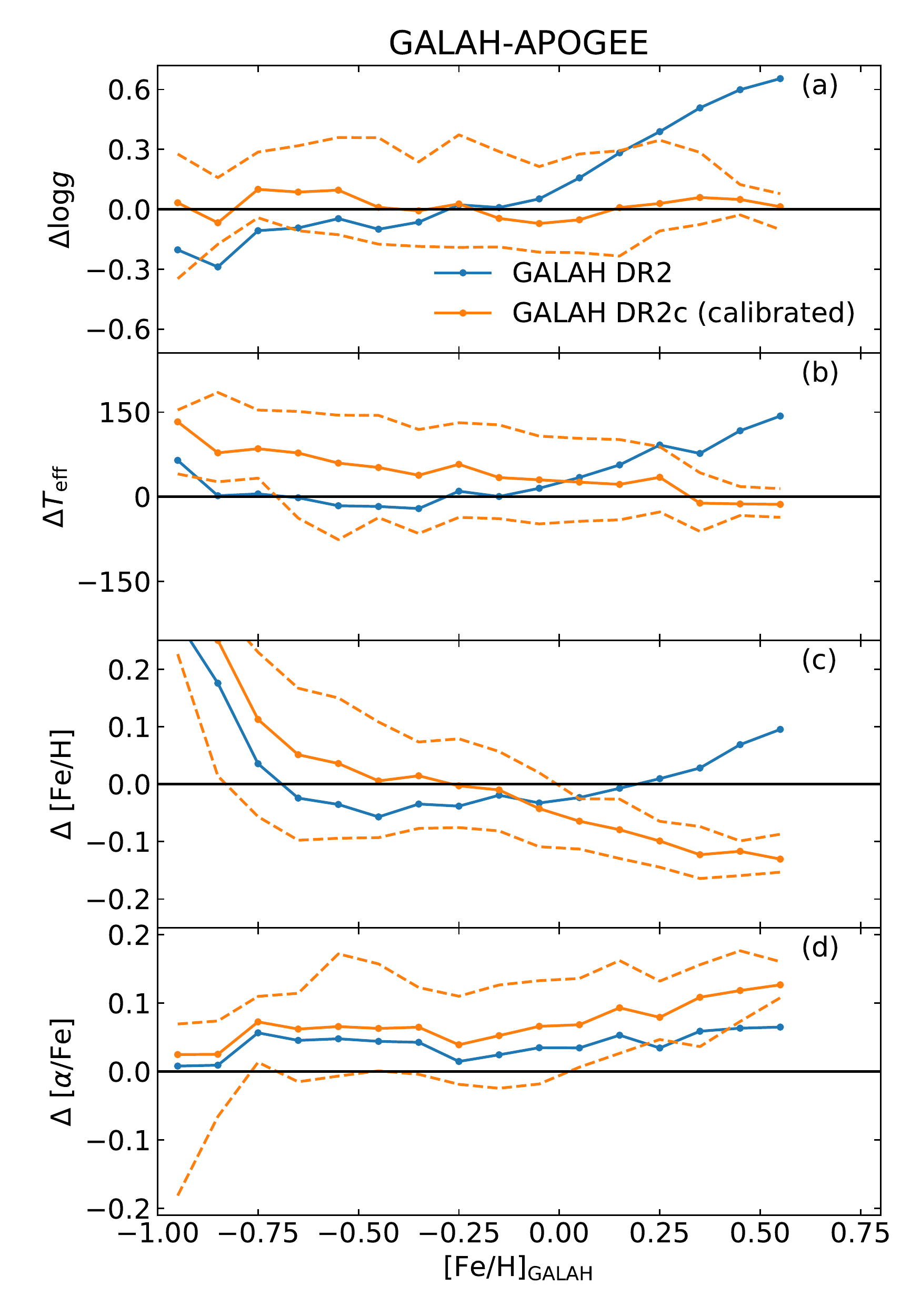}
\caption{Comparison of GALAH DR2 stellar parameters with APOGEE stellar parameters. Results corresponding to both uncalibrated and calibrated GALAH DR2 data are shown.
\label{fig:galah_vs_apogee}}
\end{figure}
\begin{table}[htb!]
\caption{Polynomial coefficients of calibration equation
$y_{\rm calib}=y+c_0+c_{1}{\rm [Fe/H]}+c_{2}{\rm [Fe/H]}^2$
to correct for systematics in the Cannon based estimates against the SME based estimates. The equation was derived using giants having $\nu_{\rm max}$ estimates from asteroseismology and with $-1.5<{\rm [Fe/H]}<0.3$. The calibration is applied to giants with ${[\rm Fe/H]}>-1.5$, the giants are identified using the  \citet{2011AJ....141..108C} definition.}
\begin{tabular}{@{}llll}
\hline
$y$  & $c_2$ & $c_1$ & $c_0$ \\
\hline
$\log g$ & +3.4987e-01 & +7.4591e-01 & +1.5727e-01 \\
$T_{\rm eff}$ K & +1.5658e+02 & +2.1861e+02 & -3.9895e+00  \\
${\rm [Fe/H]}$ & +1.9087e-01 & +2.7875e-01 & +2.4761e-02 \\
$[\alpha/{\rm Fe}]$ & -2.5775e-02 & -4.1510e-02 & -3.2592e-02 \\
\hline
\end{tabular}
\label{tab:cannon_sme_poly}
\end{table}
\begin{figure*}[htb!]
\centering \includegraphics[width=1.0\textwidth]{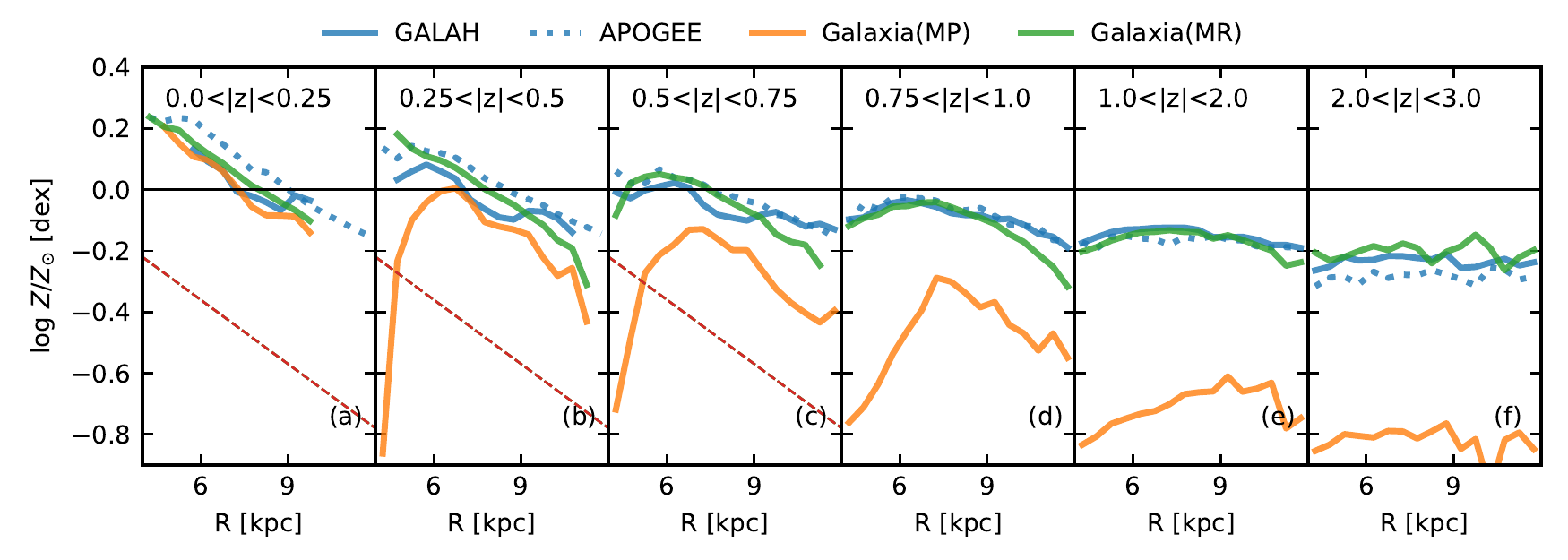}
\caption{Mean metallicity as function of Galactocentric radius $R$ for different slices in height $|z|$. The observed results are from GALAH (calibrated) and APOGEE.  Selection-function-matched {\sl Galaxia} predictions based on two different Milky Way models (the old MP model and the new MR model) are also shown. The metallicity profile has a gradient close to the plane but is flat above the plane. The dashed line for reference denotes [M/H] with a radial gradient of -0.07 dex/kpc.
\label{fig:feh_vs_R}}
\end{figure*}

In \autoref{tab:al_feh}, for GALAH two different estimates are given, the first is based on the GALAH DR2 pipeline, the second named GALAH DR2c is based on a calibration correction that we derive and apply to the GALAH DR2 estimates. GALAH DR2 estimates are based on {\it The Cannon} method \citep{2015ApJ...808...16N}, which was trained on results from the SME \citep{2017A&A...597A..16P} pipeline . However, as shown in \autoref{fig:galah_vs_seism}, for giants we find subtle systematics in the GALAH DR2 stellar parameters  compared to that of SME estimates, where $\nu_{\rm max}$ estimated from asteroseismology was used as a prior. The systematics are particularly significant for stars with [Fe/H]> 0.0.  We use the seismic giants in the SME training set to recalibrate the GALAH results. The coefficients of the calibration equation are given in \autoref{tab:cannon_sme_poly}. A comparison of GALAH stellar parameters (both calibrated and uncalibrated) with those from APOGEE for common stars is shown in \autoref{fig:galah_vs_apogee}. The calibrated GALAH gravities and temperatures match better with APOGEE. In the range $-0.5<{\rm [Fe/H]}<0.0$, where the majority of the sample is found, the calibrated [Fe/H] also matches better with APOGGE. Outside this range some systematics exist. The [$\alpha$/Fe] shows slight offsets in zero points but no significant trend is seen.

We see from \autoref{tab:al_feh} that the estimates for the mean metallicity of stars above the midplane from APOGEE, the GALAH DR2, and the GALAH DR2c agree to within 0.07 dex. The lowest values are for GALAH DR2c and the highest are for GALAH DR2. We now investigate if the metallicity of stars in $5<R/{\rm kpc}<11$ and $1<|z|/{\rm kpc}<2$ is really representative of the thick disc metallicity. We show in \autoref{tab:al_feh} estimates from a mock {\sl Galaxia} sample matched to the GALAH survey, with a thick disc having a mean
[M/H] metallicity of -0.18, for stars lying in the same spatial selection. The estimated metallicity is higher by only 0.01 dex compared to the metallicity of the thick disc that was used in the model. This suggests that the metallicity of stars with $5<R/{\rm kpc}<11$ and $1<|z|/{\rm kpc}<2$ is indeed close to the actual metallicity of the thick disc but is probably higher by 0.01 dex. Note, GALAH and APOGEE are magnitude limited surveys, so their samples are not volume complete and this can bias the estimates of the mean metallicity.

We now measure the metallicity of the thick disc more accurately by taking the selection function into account. For this, we fitted a Galactic model to the GALAH DR2c data lying within $5<R{\rm kpc}<11$, $0.75<|z|/{\rm kpc}<3$, $12<V_{JK}<14$, and with $\texttt{field\_id}< 6546$, and having positive rotation, using our importance-sampling framework (\autoref{sec:impsamp}).
\begin{figure*}[htb!]
\centering \includegraphics[width=0.95\textwidth]{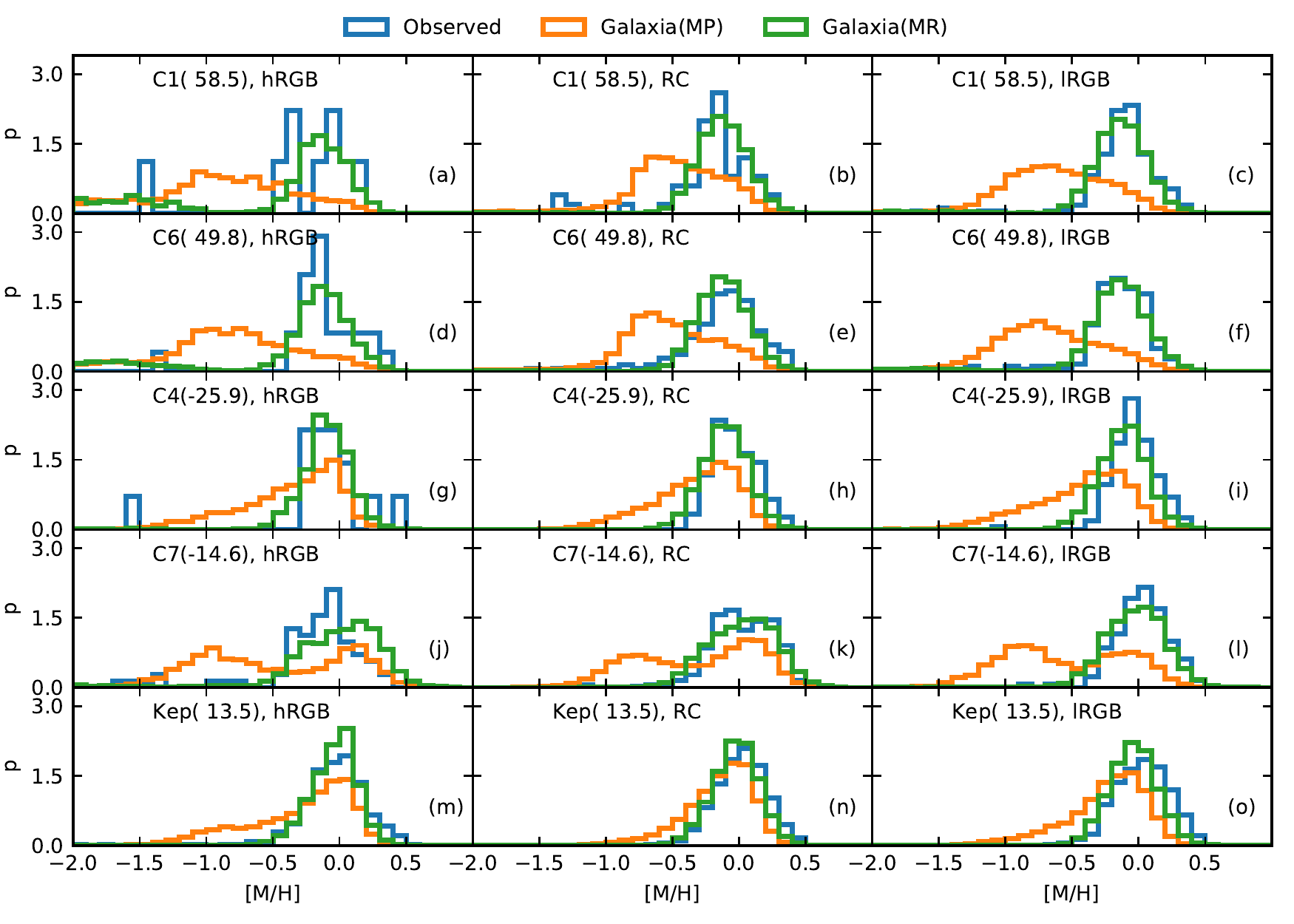}
\caption{The distribution of metallicity[M/H] for RGB and red clump stars with seismic detections from K2 campaigns C1, C4, C6, C7, and {\it Kepler}.
The left panels (a,d,g,j,m) show high luminosity RGB stars, middle panels (b,e,h,k,n) show red clump stars, and right panels (c,f,i,l,o) show low luminosity RGB stars. Observed data is compared against predictions from
theoretical models, the default model of {\sl Galaxia}-MP, which has a metal poor thick disc and the new model MR, which has a more metal rich thick disc.  The metallicity for the K2 stars is from the K2-HERMES survey while for the {\it Kepler} stars we adopt APOGEE-DR14 metallicities. The Galactic latitude of each field is enclosed in parenthesis.
\label{fig:meh}}
\end{figure*}
\begin{figure*}[htb!]
\centering \includegraphics[width=0.95\textwidth]{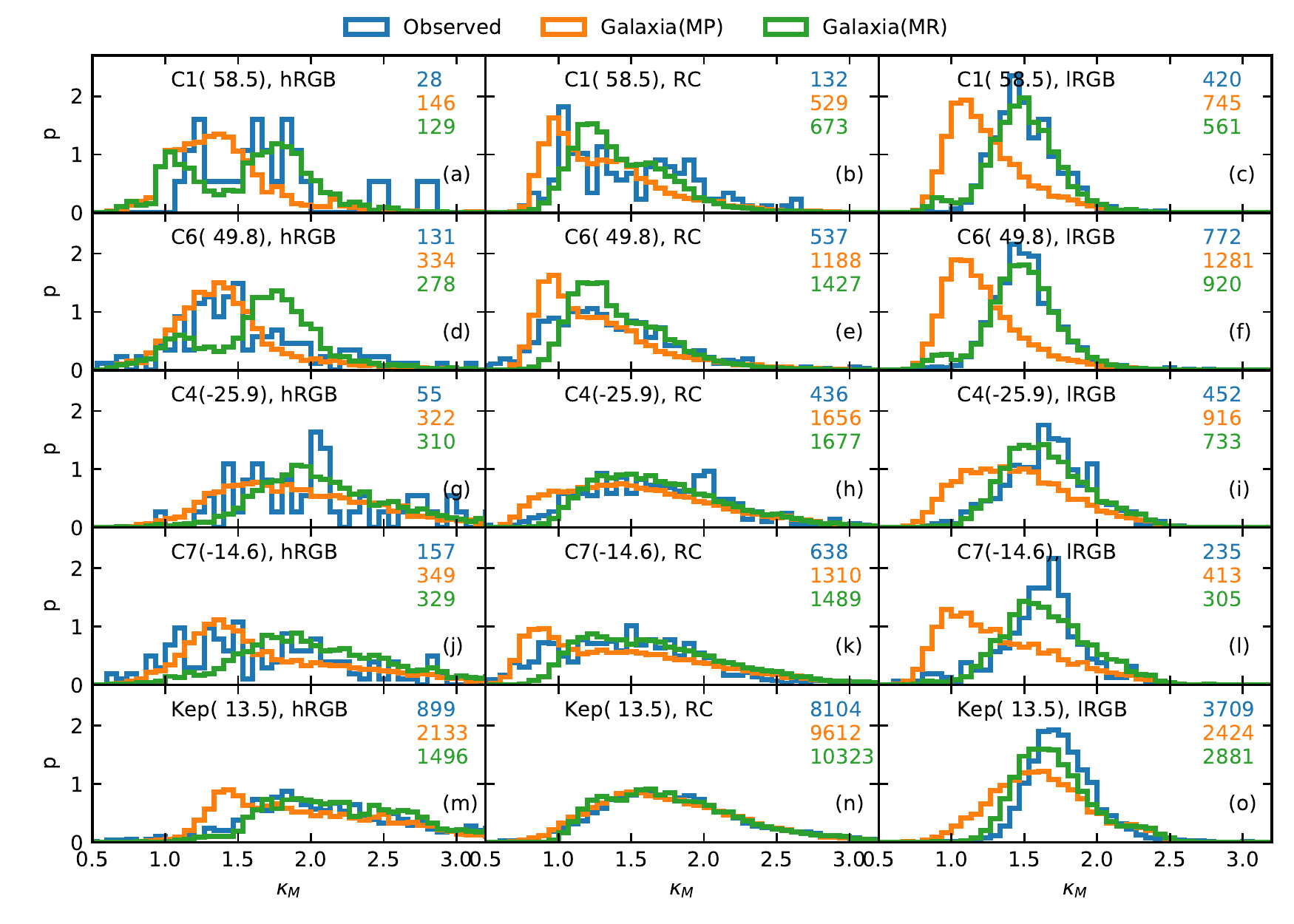}
\caption{The distribution of $\kappa_M$ for RGB and red clump stars for campaigns C1, C4, C6, C7 and {\it Kepler}.
The annotation and order of the panels is the same as in \autoref{fig:meh}. For each panel, the number of stars in each sample are listed on the right hand side.
\label{fig:kappa_m}}
\end{figure*}

The fitting procedure gave a mean metallicity of $-0.16$ with a spread of 0.17 for the thick disc and a mean metallicity of 0.0 for the oldest three thin disc subpopulations.
While fitting the model, the spread of the thin disc metallicity with age, and the metallicity of the youngest four thin disc subpopulations  was left unchanged as in the Besan\c{c}on model \citep{2003A&A...409..523R}. We also assumed that the three oldest thin disc subpopulations (age 3 to 10 Gyr) have the same mean metallicity.
We make these assumptions because
we do not make use of the age information in our fitting, and without ages it is difficult to constrain the age metallcity relation. In our fitting, the limited ability to constrain the age metallcity relation comes from the fact that the vertical height of a star is correlated with age.
Additionally, the age of the thick disc was assumed to span from 9 to 11 Gyr. We also checked with an age span of 10 to 12 Gyr for the thick disc and found that it gives the same best fit parameters.
The slight decrease in the mean metallicity of the thick disc compared to the estimate based on simply measuring the metallicity of stars within $5<R/{\rm kpc}<11$ and $1<|z|/{\rm kpc}<2$, is due to the inclusion of stars lying between $2<|z|/{\rm kpc}<3$ in the fitting process. This suggests that there is a small vertical gradient in the metallicity of the thick disk as found by others previously \citep{2011A&A...535A.107K, 2018MNRAS.476.5216D}.
Note, stars in $2<|z|/{\rm kpc}<3$
were not included in the former scheme as they could be contaminated with stars from the metal-poor stellar halo and could suffer from volume incompleteness. However, they are included in the later scheme because it takes both these effects into account.

In \autoref{fig:feh_vs_R} we show the mean metallicity as function of Galactocentric radius $R$ for different slices in height $|z|$. Results from APOGEE and GALAH DR2c are shown separately. We also plot {\sl Galaxia} predictions from the MP and MR models. To eliminate stars belonging to the halo we restrict the analysis to stars having positive rotation about the Galaxy.
The MP model clearly has a thick disc, which is too metal poor to fit the observed data for $|z|>1$ kpc. The new MR model is a very good fit to the GALAH data. It also fits the APOGEE data very well, except for the slice closest to the midplane and the slice furthest from the midplane. Compared to GALAH, stars in APOGEE are metal rich close to the midplane, but they progressively become metal poor with increasing height above the midplane. These differences could be due to systematics in abundances between the two surveys, but could also be due to the different selection function of the surveys. The effect of selection function is clearly visible in \autoref{fig:feh_vs_R}f corresponding to the top most slice ($2<|z|/{\rm kpc}<3$). Here, the MR model has a thick disc with a mean metallicity of -0.16. However, the mean metallicity of stars in this slice is below -0.2. This is because metal poor stars are more luminous and are visible furthest in a magnitude limited survey.
\autoref{fig:feh_vs_R} shows that close to the midplane there is a strong radial metallicity gradient. As we move away from the midplane the gradient diminishes progressively to zero. Close to the plane, a radial gradient of -0.07 (dashed line), as used in the Besan\c{c}on (MP) and MR models, is roughly consistent with the observed data.

We now study the distribution of metallicity for oscillating
giants detected by K2 for which we also have HERMES spectra.
In \autoref{fig:meh} we show results separately for different campaigns (C1, C6, C4, C7, and {\it Kepler} (Kep)) and different seismic classes (hRGB, RC, lRGB). Predictions from {\sl Galaxia}-MP (orange) and the new {\sl Galaxia}-MR (green) are shown alongside the observed data (blue).
The {\sl Galaxia}-MP samples have many more metal poor stars with [M/H]$<-0.5$ than the observed stars. In some panels a double peaked distribution can also be seen
with one of the peaks being at -0.78, corresponding to the metallicity of the thick disk in the model.
The new model {\sl Galaxia}-MR samples, which has a metal rich thick disc ($\langle {\rm [M/H]} \rangle \sim -0.16$), does not show a bimodal behavior and its distribution matches very well with observations. However, slight mismatches can be seen for low latitudes fields. For RC and lRGB stars in C4 and the {\it Kepler} field the {\sl Galaxia}-MR samples are still too metal poor. For hRGB in C7 the {\sl Galaxia}-MR samples are too metal rich.

\subsection{Constraints from asteroseismology}
In this section we present results making use of the asteroseismic data. We first compare the observed distribution of seismic masses against the predictions of fiducial Galactic models. Next, we restrict our analysis to thick disc stars and assuming reasonable priors on the thick disc parameters, we demonstrate that the asteroseismic scaling relations are fairly accurate. Finally, assuming the scaling relations to be correct we estimate the age of the thick disc.

\subsubsection{Comparing observed distribution of seismic masses against predictions from Galactic models}
In \autoref{fig:kappa_m}, we study the distribution of $\kappa_M$. The order of the panels is the same as in  \autoref{fig:meh}. For hRGB stars, the overall sample size is too small to assess the quality of how well the models match the data. Both models seem to perform equally well. However, for the large hRGB {\it Kepler} sample we do see that the new model provides a visibly better match.
Now turning to the RC stars, we see across the board that the new MR model performs better than the old MP model, which predicts too many stars with $\kappa_M< 1$. Finally, for lRGB stars the MR model is significantly better than the MP model, which predicts too  many stars with $\kappa_M< 1.25$.

Having seen the qualitative trends, we now move on
to do a quantitative comparison of the observed distributions
with predictions from {\sl Galaxia}. Specifically, we want to
answer the following questions: a) does the MR model match the K2 data better than the MP model does, b) and does the MR model also provide a better match to the {\it Kepler} data, which had issues with the selection function.

\begin{table*}[tb!]\centering
\caption{Ratio of observed (CAN pipeline) median $\kappa_M$ to that predicted by {\sl Galaxia} for different giant classes. Results for two different Galactic models MP (metal poor) and MR (metal rich) are shown. Uncertainties on the computed ratio are also listed.
}
\begin{tabular}{l|ll|ll|ll}
\hline
& \multicolumn{2}{c}{hRGB} & \multicolumn{2}{|c|}{RC} & \multicolumn{2}{c}{lRGB} \\
Campaign  & {\sl Galaxia}(MP) & {\sl Galaxia}(MR) & {\sl Galaxia}(MP) & {\sl Galaxia}(MR) & {\sl Galaxia}(MP) & {\sl Galaxia}(MR) \\
\hline
1          & $1.23 \pm 0.05$ & $1.01 \pm 0.04$ & $1.15 \pm 0.03$ & $1.05 \pm 0.03$ & $1.242 \pm 0.009$ & $0.992 \pm 0.007$ \\
6          & $1.07 \pm 0.03$ & $0.87 \pm 0.02$ & $1.11 \pm 0.02$ & $0.97 \pm 0.01$ & $1.287 \pm 0.007$ & $1.002 \pm 0.005$ \\
4          & $1.07 \pm 0.04$ & $0.98 \pm 0.04$ & $1.11 \pm 0.02$ & $1.01 \pm 0.01$ & $1.18 \pm 0.01$ & $1.027 \pm 0.009$ \\
7          & $1.05 \pm 0.03$ & $0.83 \pm 0.03$ & $1.07 \pm 0.02$ & $0.92 \pm 0.01$ & $1.3 \pm 0.01$ & $1 \pm 0.01$ \\
{\it Kepler}     & $1.1 \pm 0.01$ & $0.96 \pm 0.01$ & $1.021 \pm 0.003$ & $1.009 \pm 0.003$ & $1.086 \pm 0.003$ & $1.037 \pm 0.002$ \\
\hline
\end{tabular}
\label{tab:kappa_m}
\end{table*}

The $\kappa_M$ distributions are in general unimodal. At the most basic level a unimodal distribution over a finite domain can be characterized by a median. We first estimate the medians and then compute the ratio of medians  between the observed and predicted  distributions, which we show in \autoref{tab:kappa_m}. Ideally, we expect the ratio to be close to one, but in previous work based on {\it Kepler} data, we found the median ratio to be larger than one (1.06).

The new MR model, anchored on GALAH metilicities of the thick disc, is undoubtedly better than the old MP model. For almost all giant classes and campaigns, the median ratio for the new MR model is closer to unity than for the old MP model. The only two exceptions are hRGB for C6 and C7, where the ratio is about 0.85, i.e., the model overpredicts the masses. However, these samples suffer from low number statistics. Additionally
for the hRGB stars in C7, we also noticed that the MR model overpredicts the metallicity \autoref{fig:meh}, and this will lead to overestimation of masses in the MR model.

\subsubsection{Testing the accuracy of the asteroseismic mass scaling relation}
\begin{figure}
\centering \includegraphics[width=0.48\textwidth]{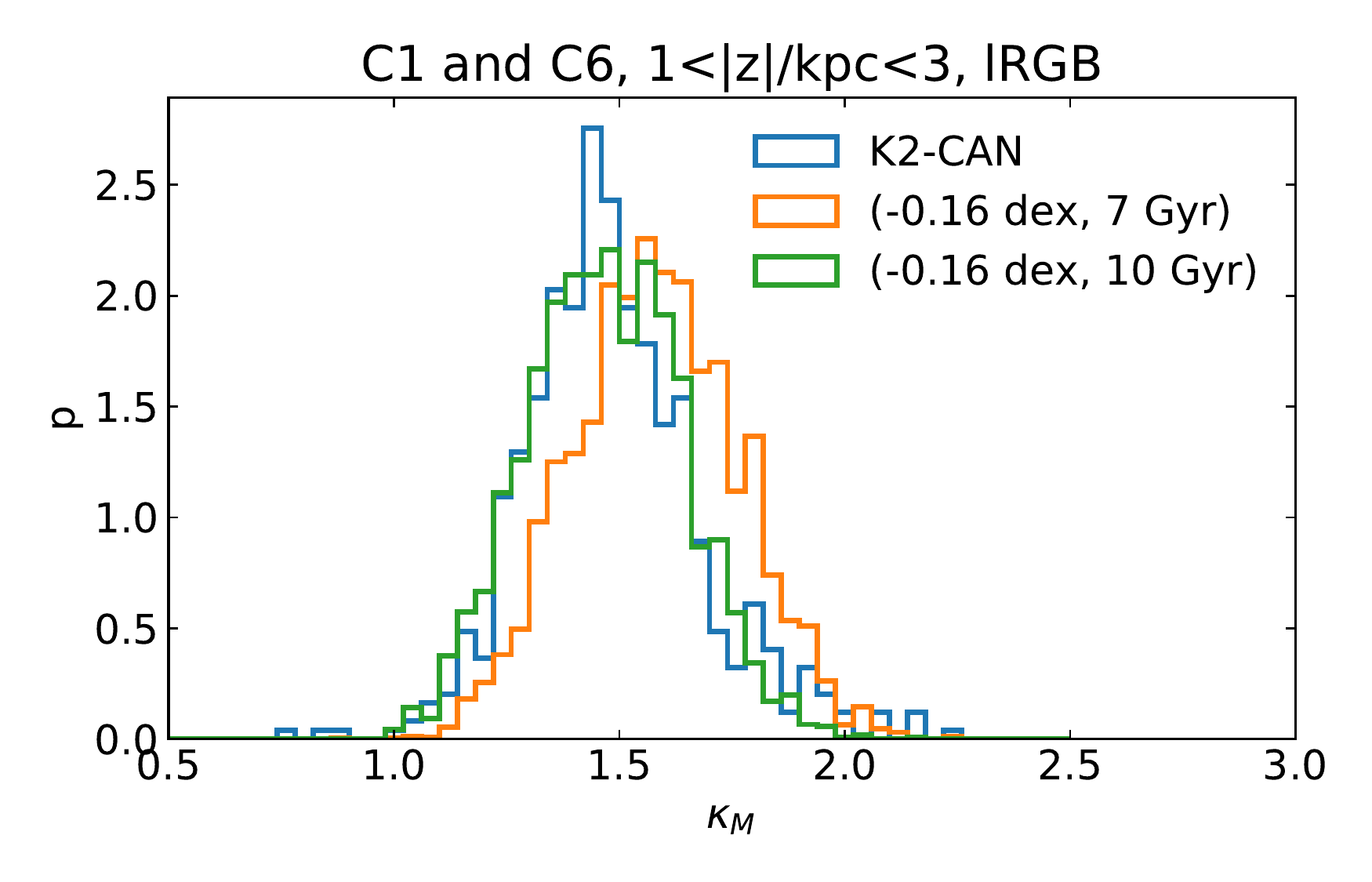}
\caption{The distribution of $\kappa_M$  for lRGB stars in K2 campaigns C1 and C6 that lie between $1<|z|/{\rm kpc}<3$. Shown alongside are mass distributions corresponding to   simple stellar populations with a Gaussian metallicity distribution and a uniform age distribution (with a width of 2 Gyr). The mean metallicity  and the mean age of each stellar population is given in the legend.
\label{fig:can_c16_kappa_m}}
\end{figure}

\begin{figure}
\centering \includegraphics[width=0.48\textwidth]{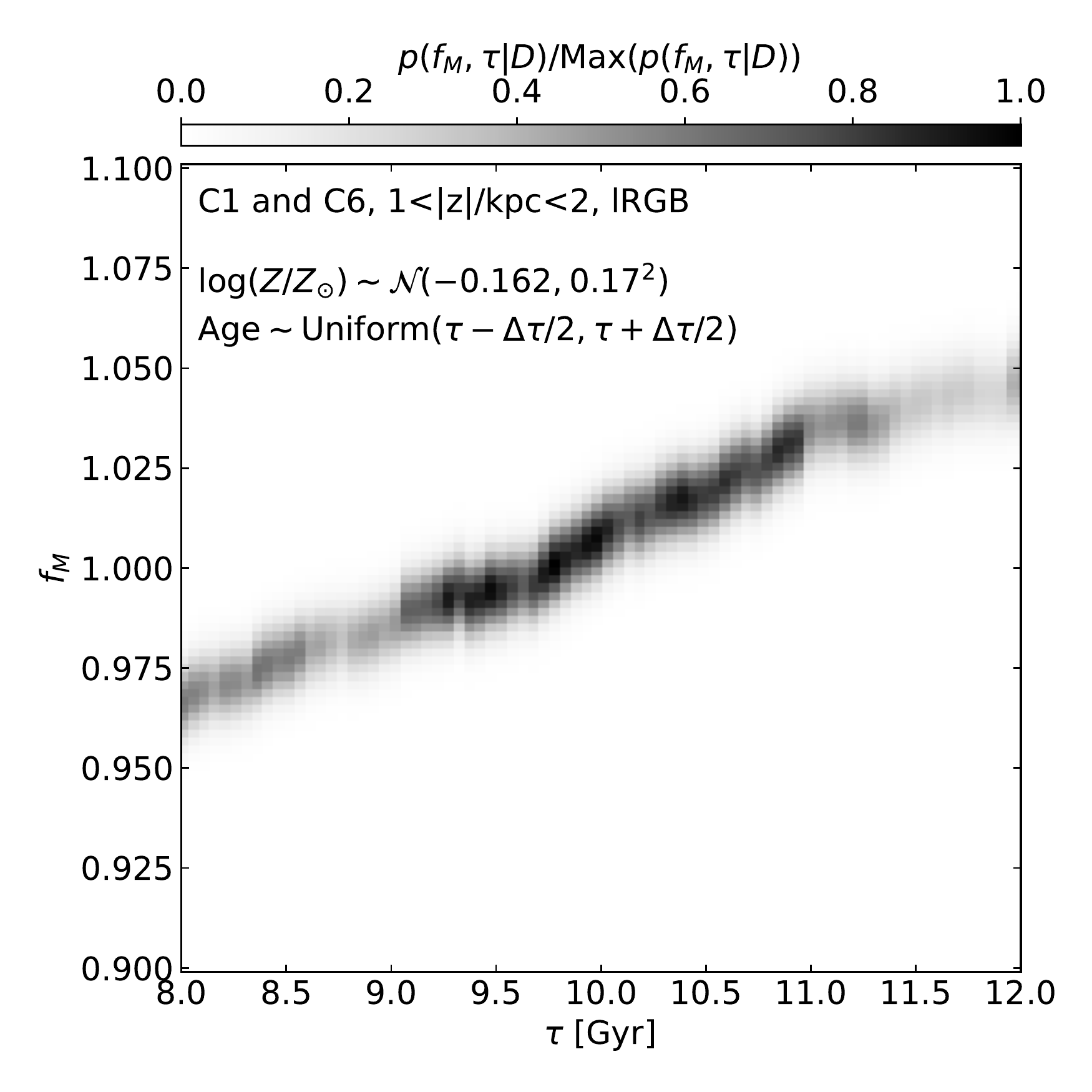}
\caption{The posterior distribution of $f_{M}$ and mean age of the thick disc $\tau$ obtained using  lRGB stars in K2 campaigns C1 and C6 that lie between $1<|z|/{\rm kpc}<3$. The width $\Delta \tau$ of the age distribution  was assumed to be 2 Gyr.
\label{fig:likelihood_2}}
\end{figure}

The fact that the mass distribution
of the new model MR matches the observed seismic masses so well, suggests that the asteroseismic scaling relations are fairly accurate. In the following we will explore this more quantitatively by limiting the analysis to a single Galactic component and imposing reasonable non-seismic priors on its parameters.
To do this, we study the mass distribution of stars lying between $1<|z|/{\rm kpc}<3$.
The Galactic model predicts that about 90\% of these stars should be thick disc stars, so we can model them as a simple stellar population characterized by some age distribution and metallicity distribution. We have already shown that the metallicity distribution of this population can be represented by $\mathcal{N}(-0.16,0.17^2)$.
In the following we present several pieces of evidence suggesting that the mean age of this high $|z|$ population should be between 8 to 12 Gyr. 
Firstly, \autoref{fig:alpha_feh_dist} shows that stars between $1<|z|/{\rm kpc}<2$ are enhanced in $\alpha$ element abundances and form a distinct sequence in the abundance space.
Using dwarf and subgiants in the solar neighborhood, it has been shown  that the stars in the $\alpha$-enhanced sequence are typically older than 10 Gyr  (\citet{Bensby2014} Figure 22 and \citet{2017A&A...608L...1H} Figure 3).
Secondly, chemical evolution models predict that $\alpha$-enhanced stars must have formed within the first 1 Gyr of the star formation history of the Milky Way, or else the contribution from Type-Ia supernovae would have introduced too much iron and hence brought the value of [$\alpha$/Fe] down \citep{pagel_book_2009}. When the above fact is
combined with Figure 3 from   \citet{2017A&A...608L...1H}, which suggest that the oldest thin disc stars (stars not enhanced in [$\alpha$/Fe]) are around 8 to 10 Gyr old, we reach the conclusion that the
$\alpha$-enhanced population must be
older than 8 to 10 Gyr. Finally, \citet{2017ApJ...837..162K} provide one of the most precise and accurate estimates on the mean age of the thick disc using nearby white dwarfs. They estimate the mean thick disc age to be between 9.5 to 9.9 Gyr, with a random uncertainty of about 0.2 Gyr. 
Hence, based on these observational evidence, a reasonable prior for the mean age of the thick disc is 8 to 12 Gyr.

To test the asteroseismic mass scaling relation we select the
lRGB stars in K2 campaigns C1 and C6 that lie between $1<|z|/{\rm kpc}<3$.
We avoid campaigns C4 and C7
because they point into the Galactic plane and hence lack high $|z|$ stars.  We restrict our test to
lRGB stars because for these stars there is almost 100\% probability both to detect $\nu_{\rm max}$ and to detect $\Delta \nu$ when a $\nu_{\rm max}$ has been measured.
The distribution of $\kappa_M$ for the lRGB stars is shown in
\autoref{fig:can_c16_kappa_m}. The distributions of $\kappa_M$ for a stellar population with a metallicity distribution of $\mathcal{N}(-0.16,0.17^2)$
and a mean age of $\tau=10$  Gyr
is also shown alongside, showing a good match to the observed distribution.
However, the distribution for the stellar population with
$\tau=7$ Gyr but the same metallicity distribution as before, is shifted too far to the right.
Now, to quantify the accuracy of the asteroseismic mass scaling relation (Eq.~\ref{equ:scaling_m}), we introduce a factor $f_{M}$, that is multiplied to $\kappa_M$ for stars in the model to get a `corrected' mass, and then we investigate how close to unity this correction factor is when enforcing that the observed and model mass distributions match.

The posterior distribution of $f_{M}$ and the age, $\tau$, conditional on our data $D$ is given in \autoref{fig:likelihood_2}.
For the mean age of the high $|z|$ population we assume a flat prior in the range 8 to 12 Gyr.
The analysis was done using the importance sampling framework discussed in \autoref{sec:impsamp} and taking the photometric selection function into account. The figure shows that $f_{M}$ depends upon $\tau$ and  varies between 0.97 and 1.05 for the adopted range of $\tau$. This would translate into a maximum deviation of the \numax\ scaling relation (Eq.~\ref{equ:scaling_numax}) of 1-2\% if the \dnu\ scaling relation (Eq.~\ref{equ:scaling_dnu}) is true. Or alternatively, that the maximum deviation of the \dnu\ scaling relation would be about 1\% if the \numax\ scaling relation is true.  Now, if both the \dnu\ and the \numax\ relations are incorrect but conspire to cancel out their inaccuracy when using the mass scaling relation (Eq.~\ref{equ:scaling_m}), one could in principle have a scenario where large deviations of the \dnu\ and \numax\ relations could be hidden in our mass test. However, this seems not to be the case because when testing the radius scaling relation
\be
\frac{R}{\rm R_{\odot}}=\left(\frac{\nu_{\rm max}}{\nu_{\rm max,\odot}}\right)\left(\frac{\Delta \nu_{\odot}}{\Delta \nu_{\odot}}\right)^{-2}\left(\frac{T_{\rm eff}}{T_{\rm eff,\odot}}\right)^{0.5},
\ee
which is based on different powers of \dnu\ and \numax, Zinn et al (submitted) finds agreement between seismic and Gaia radii at the 1\% level. Hence, in combination these mass and radius scaling relation tests show strong evidence that the individual \dnu\ and \numax\ scaling relations that go into the mass and radius scaling relations are in fact astonishingly accurate.

\subsubsection{Constraining the age of the thick disc}
\begin{figure}[tb]
\centering \includegraphics[width=0.5\textwidth]{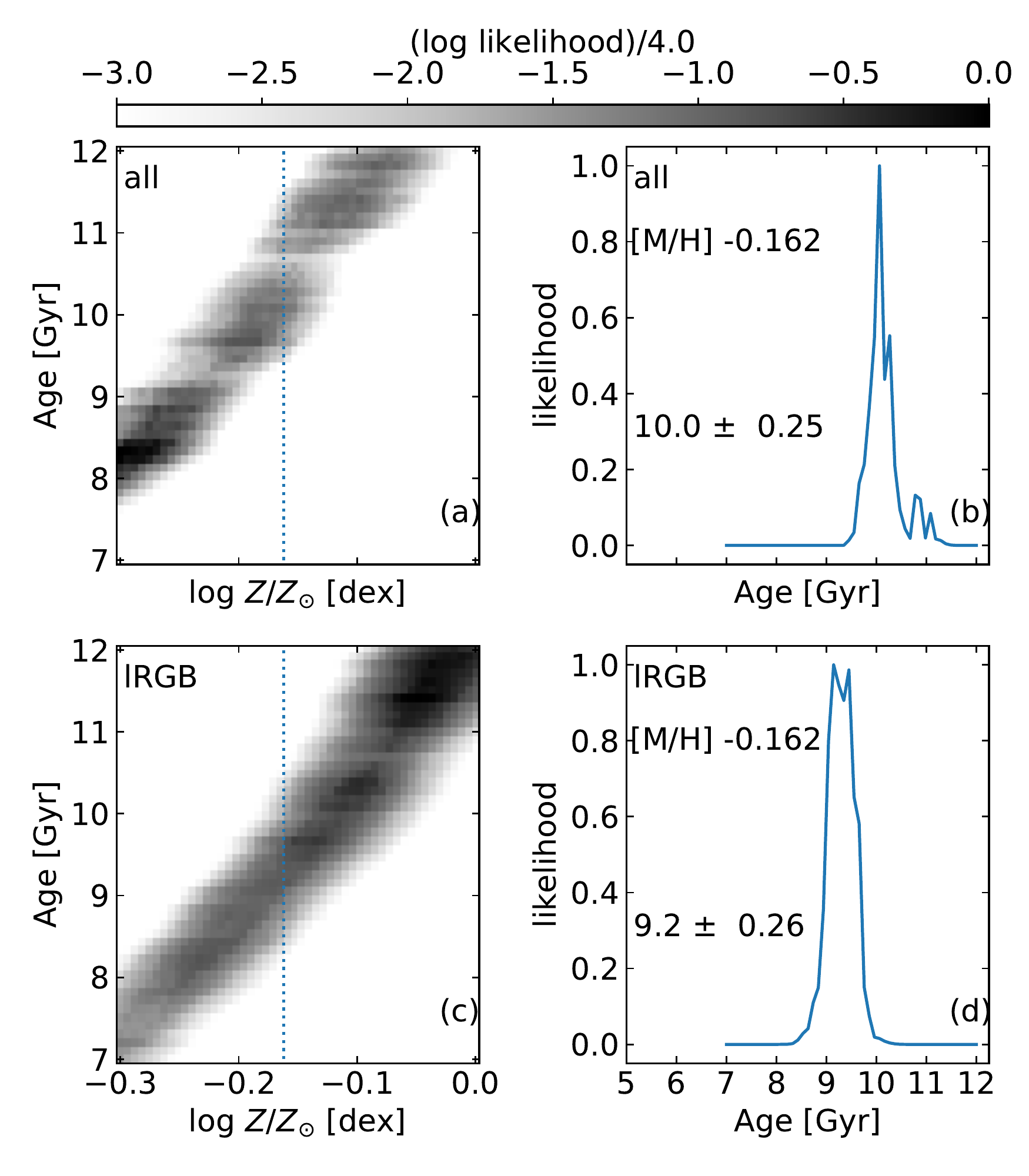}
\caption{(a,c) Likelihood of age and metallicity of the thick disc using asteroseismic information from K2 campaigns C1, C4, C6, and C7. (b-d) The likelihood of thick disc age
assuming the thick disc metallicity to be $\log (Z/Z_{\odot})=-0.162$, as estimated using the GALAH survey in \autoref{tab:tb3}. In the top panels, the likelihood is computed using all oscillating giants, while in the bottom panels, only low luminosity giants (\autoref{sec:giant_class}) are used.
\label{fig:likelihood}}
\end{figure}

Having established that the asteroseismic scaling relations are good to a high degree of accuracy, it would seem reasonable to now turn the problem around. Hence, in the following we assume the relations to be true and use the observed values of $\kappa_M$ to estimate the age and metallicity of the thick disc. We do this using the importance sampling framework discussed in \autoref{sec:impsamp}. Here, we use the FL Galactic model from \autoref{tab:gmodels} as the base model and reweight it to simulate samples corresponding to different values of the  parameters of the model.
We compute the likelihood of the observed $\kappa_M$ values given the model for different values of the mean metallicity, $\log Z/Z_{\odot}$, and mean age for the thick disc.
Given the unknown selection function of the {\it Kepler} data, only data from the K2 campaigns were used. The results are shown in \autoref{fig:likelihood}. We adopted a duration of 2 Gyr for the star formation episode of the thick disc. We also investigated shorter (1 Gyr) and longer (3 Gyr) star formation durations and found that the results were not too sensitive to the exact choice of the duration.

\autoref{fig:likelihood}a shows the likelihood when considering all giants. \autoref{fig:likelihood}b  shows the likelihood when only lRGB giants are used. It can be seen that when we only consider the asteroseismic information, age is degenerate with metallicity. 
A decrease in the adopted metallicity by 0.1 dex can decrease the inferred age by about 2 Gyr. \autoref{fig:likelihood}a  shows that a metal poor thick disc cannot be old. For example, a thick disc with $\log Z/Z_{\odot}=-0.3$ will have an age of about 8 Gyr and would be even younger if it was  more metal poor (such as the old MP model).
For a star with a given mass, the decrease in age with a decrease in metallicity is expected because a low metallicity star evolves much faster along the HR diagram, compared to a high metallicity star.

\autoref{fig:likelihood}b,d shows the likelihood as a function of age when we fix the metallicity to -0.16 as suggested by the spectroscopic data. Using all giants we get a mean age of 10 Gyr, if only lRGB stars are used  we obtain 9.2 Gyr. Both estimates are consistent with the traditional idea of an old thick disc.

\section{Discussion and Conclusions}\label{conclusions}
Asteroseismology can provide ages for giant stars
and hence is a promising tool for studying Galactic structure and evolution. However, it has proven to be difficult to check the accuracy of the ages and masses estimated by asteroseismology, due to the shortage of independent estimates of mass and age. Population synthesis based Galactic models, provide an indirect way to validate
the asteroseismic estimates. However, previous studies
using the {\it Kepler} mission revealed that the models predict too many low mass stars as compared to observed mass distributions, raising doubts on the accuracy of the asteroseismic estimates, the Galactic models,
and/or the selection function. In this paper,
we revisit this important problem by analyzing
asteroseismic data from the K2 mission, which has a well defined selection function.
For the first time, we show that if the metallicity distribution in the Galactic models is updated to measurements from recent spectroscopic surveys, the distribution of asteroseismic masses is in good agreement with the model predictions. Using thick disc stars we show that the asteroseismic mass scaling relation for low luminosity red giants should be accurate to 5\%. This is in agreement with findings of \citet{2018MNRAS.476.3729B} who tested the seismic relations using three eclipsing binary systems.

We identify three main factors, which, if not taken into account, can lead to discrepancies between
observed asteroseismic masses and model predictions.
First, in addition to age, the mass distribution giant stars in a stellar population is very sensitive to its metallicity, hence it is important to get the metallicity distribution of the various Galactic components in a model to agree with observations. Second, certain
Galactic components are significantly enhanced
in abundance of $\alpha$ elements and this should be taken into account, either directly by using $\alpha$ enhanced isochrones, or indirectly by increasing the
effective metallicity of the solar scaled isochrones.
Third, the $\Delta {\nu}$ scaling relation is not strictly valid and there exists theoretically motivated corrections, which should be applied.
It was already shown in a previous study \citep{2016ApJ...822...15S} that the correction is such that it helps to reduce the mass discrepancy.

Using a forward modelling approach, where we take the Besan\c{c}on Galactic model as a prior, we fit for the effective metallicity $Z$ (taking $\alpha$ enhancement into account) of the thin and the thick disc using the GALAH data. We find the mean $\log Z/Z_{\odot}$ of the thin disc to be $0.0$ and that of the thick disc to
be -0.162 (with a dispersion of 0.17, see \autoref{tab:al_feh}). This is in good agreement with data from the APOGEE survey. This is a significant revision of thick disc from a value of [Fe/H]$=-0.78$ as used in the Besan\c{c}on model. An increase of about 0.14 dex in $\log Z/Z_{\odot}$ is due to taking the $\alpha$ enhancement into account, but about 0.5 dex is due to revision of [Fe/H]. For example, if we consider stars in $5<R/{\rm kpc}<7$ and $1<|z|/{\rm kpc}<2$, which mostly come from the thick disc, both GALAH and APOGEE suggest a mean [Fe/H]$\sim-0.30$ for the thick disc.

Using a forward modelling approach, we also fit for the age of the thick disc using the asteroseismic data. We find the mean age to be about $9.2 - 10 \pm 0.25$ Gyr (redshift of about 1.6), which is broadly consistent with the idea of the thick disc being old and formed early on in the history of the Galaxy.
What exactly do we mean by thick disc? Traditionally the thick disc was identified as the component with higher scale height in the solar annulus. Observations also suggest the thick disc to be  distinct in elemental abundances from the thin disc. Two sequences
$\alpha_{+}$ and $\alpha_{\rm o}$ can be seen in the ([$\alpha$/Fe],[Fe/H]) plane, with the former
(having higher [$\alpha$/Fe]) being the thick disc and the
later the thin disc. New results \citep{2011ApJ...735L..46B,2012ApJ...753..148B, 2017ApJS..232....2X,2017MNRAS.471.3057M} suggest that the scale length of the $\alpha_{+}$ sequence is shorter than that of the $\alpha_{\rm o}$ sequence. Chemical evolution models require
the $\alpha_{+}$ sequence to be old. In our forward modelling we do not identify the thick disc using elemental abundances. Instead the thick disc is indirectly identified by our prior for the spatial distribution of thin and thick disc stars. In the model, stars with $|z|>1$ kpc are dominated
by thick disc. The majority of the thick disc stars in our model come from the high latitude campaigns C1 and C6, and these stars have Galactocentric radius similar to that of the Sun. So our thick disc metallicity and age measurements are representative of the properties of the stellar population that roughly dominates in the region $|z|/{\rm kpc}>1$ and $6<R/{\rm kpc}<10$.

Our thick disc age estimate is consistent with
previous studies that estimated
the mean age independent
of asteroseismology. For example,
it is consistent with results by \citet{Bensby2003} who estimate the age to be $11.2 \pm 4.3$ using F and G dwarfs. It is consistent with
\citet{2017ApJS..232....2X} from LAMOST using main sequence turn-off stars and subgiants, where they show that stars with, $|z|>1$ kpc have a median age close to 10 Gyr and are $\alpha$ enhanced.
It is consistent with results by \citet{2017MNRAS.471.3057M}
from APOGEE using giants, where they show
that $\alpha$ enhanced stars
have significantly larger scale height and
their mean age is close to or larger than 10 Gyr. However, the age estimates in \citet{2017MNRAS.471.3057M} are anchored on the asteroseismic age scale.
Finally, our estimate ($9.2-10\pm 0.25$ Gyr) is in excellent agreement with estimates of \citet{2017ApJ...837..162K} of $9.5-9.9\pm 0.2$ Gyr using white dwarfs, an estimate that is very accurate and independent of both asteroseismology and the isochrones.

Although we find that the observed mass distributions are in good agreement with predictions by Galactic models, some small unexplained differences do remain. For lRGB, the predicted mean of the mass distributions for K2 campaign C4 and {\it Kepler} are higher by about 3\%. We also see differences in metallicity
distributions for these samples and this could potentially be responsible for the mass differences.
For hRGB and red clumps, the mean predicted mass is lower than observed, for campaigns
C6 and C7. This could be due to imperfections in the model, but could also be related to the
fact that the detection of $\Delta \nu$
is not complete for these stars.

We present the selection function for four K2 campaigns and discuss detection biases associated with the K2 data, which should be taken into account when using the K2 data. Probability to detect  $\nu_{\rm max}$ varies with both $\nu_{\rm max}$ and apparent magnitude.
Low-luminosity stars have lower
oscillation amplitudes and cannot be detected at fainter magnitudes. Even after we account for the effect of oscillation amplitude and apparent magnitude, comparison with Galactic models show that the overall detection rate for $\nu_{\rm max}$ is about 72\%. Using a deep-learning-based
pipeline improves the detection rate to 78\%, which is still quite low. It is not yet clear as to why the detection rate is low.
It could be that certain specific type of stars (e.g., red clumps or metal poor stars) have lower than expected oscillation amplitudes, or it could be an unknown instrumental effect, or even a problem with the Galactic model. There are also biases related to  detecting $\Delta \nu$ in the K2 data.
The probability to detect $\Delta \nu$ has a strong dependence on $\nu_{\rm max}$, it is less than 1 for $\nu_{\rm max}<50\ \mu$Hz, but is otherwise close to 1.  Significant campaign to campaign differences are also seen, which needs further investigation. To take the detection biases into account, we propose to split up the stars into different giant classes based on their detection probabilities.
Asteroseismic pipelines also show small systematic offsets in estimation of $\nu_{\rm max}$ which need further investigation.

Using the seismic sample, we find that the stellar parameters for giants in GALAH DR2, which are based on the data-driven {\it The Cannon} scheme, have systematic differences with respect to estimates based on the model-driven SME scheme that is anchored to seismic $\nu_{\rm max}$ values. Differences are most significant for stars with [Fe/H]>0. We provide analytical functions to correct for them. The reason for the systematic offsets is because the giants in the training set used by {\it The Cannon} were dominated by non seismic giants. In the absence of seismic $\nu_{\rm max}$, the SME gives biased results. SME with Gaia DR2 parallaxes as prior alleviates this problem, however,  Gaia DR2 parallaxes were not available at the time of publication of GALAH DR2.

In near future, we will have a much larger sample of stars with asteroseismology from both the K2 and the TESS \citep{2015ApJ...809...77S} missions. This will  allow us to fit more detailed models of our Galaxy than done here. Specifically, we can study the
properties of the stellar populations as a function of age with much finer age resolution.
Future, spectroscopic surveys, such as the, second phase of GALAH, 4MOST \citep{2016SPIE.9908E..1OD}, WEAVE \citep{2018SPIE10702E..1BD}, and SDSSV \citep{2017arXiv171103234K}, will
also produce large samples of stars with age
estimates purely from spectroscopy,
based on main sequence turnoff and subgiant stars or based on giants making use of the
age information encoded in carbon and nitrogen abundances. Asteroseismology in this regard is going to play a crucial role by providing independent age estimates.

\acknowledgments
S.S. is funded by University of Sydney
Senior Fellowship made possible by the office of the Deputy Vice Chancellor of Research, and partial funding from Bland-Hawthorn's Laureate Fellowship from the Australian Research Council.
The GALAH Survey is supported
by the Australian Research Council Centre of Excellence for All Sky Astrophysics in 3 Dimensions (ASTRO 3D),through project
number CE170100013.
D.S. is the recipient of an Australian Research Council Future Fellowship (project number FT1400147).
JBH is supported by an ARC Australian Laureate Fel-
lowship (FL140100278).   MJH is sup-
ported by an ASTRO-3D Fellowship.
S.B. and K.L. acknowledge funds from the Alexander von Humboldt Foundation in the framework of the Sofja Kovalevskaja Award endowed by the Federal Ministry of Education and Research. K.L. acknowledges funds from the Swedish Research Council (Grant nr. 2015-00415\_3) and Marie Sklodowska Curie Actions (Cofund Project INCA 600398).  T.Z. acknowledges financial support from the Slovenian Research Agency (research core funding No. P1-0188).  DMN was supported by the Allan C. and Dorothy H. Davis Fellowship. J.~Z. acknowledges support from NASA grants 80NSSC18K0391 and NNX17AJ40G.
\facilities{AAT}
\software{Numpy, Matplotlib}
\bibliographystyle{yahapj}
\bibliography{main}

\end{document}